\documentstyle[12pt]{article}   

\makeatletter
\@addtoreset{equation}{section}
\makeatother

\newtheorem{Definition}{Definition}[section]
\newtheorem{Lemma}{Lemma}[section]

\def\be{\begin{equation}}
\def\ee{\end{equation}}
\def\ba{\begin{eqnarray}}
\def\ea{\end{eqnarray}}
\def\a{{\cal A}}
\def\ab{\overline{\a}}

\def\Nl{{\mathchoice
{\setbox0=\hbox{$\displaystyle\rm N$}\hbox{\hbox to0pt
{\kern0.4\wd0\vrule height0.9\ht0\hss}\box0}}
{\setbox0=\hbox{$\textstyle\rm N$}\hbox{\hbox to0pt
{\kern0.4\wd0\vrule height0.9\ht0\hss}\box0}}
{\setbox0=\hbox{$\scriptstyle\rm N$}\hbox{\hbox to0pt
{\kern0.4\wd0\vrule height0.9\ht0\hss}\box0}}
{\setbox0=\hbox{$\scriptscriptstyle\rm N$}\hbox{\hbox to0pt
{\kern0.4\wd0\vrule height0.9\ht0\hss}\box0}}}}
\def\Zl{{\mathchoice
{\setbox0=\hbox{$\displaystyle\rm Z$}\hbox{\hbox to0pt
{\kern0.4\wd0\vrule height0.9\ht0\hss}\box0}}
{\setbox0=\hbox{$\textstyle\rm Z$}\hbox{\hbox to0pt
{\kern0.4\wd0\vrule height0.9\ht0\hss}\box0}}
{\setbox0=\hbox{$\scriptstyle\rm Z$}\hbox{\hbox to0pt
{\kern0.4\wd0\vrule height0.9\ht0\hss}\box0}}
{\setbox0=\hbox{$\scriptscriptstyle\rm Z$}\hbox{\hbox to0pt
{\kern0.4\wd0\vrule height0.9\ht0\hss}\box0}}}}
\def\Ql{{\mathchoice
{\setbox0=\hbox{$\displaystyle\rm Q$}\hbox{\hbox to0pt
{\kern0.4\wd0\vrule height0.9\ht0\hss}\box0}}
{\setbox0=\hbox{$\textstyle\rm Q$}\hbox{\hbox to0pt
{\kern0.4\wd0\vrule height0.9\ht0\hss}\box0}}
{\setbox0=\hbox{$\scriptstyle\rm Q$}\hbox{\hbox to0pt
{\kern0.4\wd0\vrule height0.9\ht0\hss}\box0}}
{\setbox0=\hbox{$\scriptscriptstyle\rm Q$}\hbox{\hbox to0pt
{\kern0.4\wd0\vrule height0.9\ht0\hss}\box0}}}}
\def\Rl{{\mathchoice
{\setbox0=\hbox{$\displaystyle\rm R$}\hbox{\hbox to0pt
{\kern0.4\wd0\vrule height0.9\ht0\hss}\box0}}
{\setbox0=\hbox{$\textstyle\rm R$}\hbox{\hbox to0pt
{\kern0.4\wd0\vrule height0.9\ht0\hss}\box0}}
{\setbox0=\hbox{$\scriptstyle\rm R$}\hbox{\hbox to0pt
{\kern0.4\wd0\vrule height0.9\ht0\hss}\box0}}
{\setbox0=\hbox{$\scriptscriptstyle\rm R$}\hbox{\hbox to0pt
{\kern0.4\wd0\vrule height0.9\ht0\hss}\box0}}}}
\def\Cl{{\mathchoice
{\setbox0=\hbox{$\displaystyle\rm C$}\hbox{\hbox to0pt
{\kern0.4\wd0\vrule height0.9\ht0\hss}\box0}}
{\setbox0=\hbox{$\textstyle\rm C$}\hbox{\hbox to0pt
{\kern0.4\wd0\vrule height0.9\ht0\hss}\box0}}
{\setbox0=\hbox{$\scriptstyle\rm C$}\hbox{\hbox to0pt
{\kern0.4\wd0\vrule height0.9\ht0\hss}\box0}}
{\setbox0=\hbox{$\scriptscriptstyle\rm C$}\hbox{\hbox to0pt
{\kern0.4\wd0\vrule height0.9\ht0\hss}\box0}}}}
\def\Hl{{\mathchoice
{\setbox0=\hbox{$\displaystyle\rm H$}\hbox{\hbox to0pt
{\kern0.4\wd0\vrule height0.9\ht0\hss}\box0}}
{\setbox0=\hbox{$\textstyle\rm H$}\hbox{\hbox to0pt
{\kern0.4\wd0\vrule height0.9\ht0\hss}\box0}}
{\setbox0=\hbox{$\scriptstyle\rm H$}\hbox{\hbox to0pt
{\kern0.4\wd0\vrule height0.9\ht0\hss}\box0}}
{\setbox0=\hbox{$\scriptscriptstyle\rm H$}\hbox{\hbox to0pt
{\kern0.4\wd0\vrule height0.9\ht0\hss}\box0}}}}
\def\Ol{{\mathchoice
{\setbox0=\hbox{$\displaystyle\rm O$}\hbox{\hbox to0pt
{\kern0.4\wd0\vrule height0.9\ht0\hss}\box0}}
{\setbox0=\hbox{$\textstyle\rm O$}\hbox{\hbox to0pt
{\kern0.4\wd0\vrule height0.9\ht0\hss}\box0}}
{\setbox0=\hbox{$\scriptstyle\rm O$}\hbox{\hbox to0pt
{\kern0.4\wd0\vrule height0.9\ht0\hss}\box0}}
{\setbox0=\hbox{$\scriptscriptstyle\rm O$}\hbox{\hbox to0pt
{\kern0.4\wd0\vrule height0.9\ht0\hss}\box0}}}}
\title{Gauge Field Theory Coherent States (GCS) : I. General Properties}
\author{Thomas Thiemann\thanks{thiemann@aei-potsdam.mpg.de}\\
Albert-Einstein-Institut, Max-Planck-Institut
f\"ur Gravitationsphysik,\\ Am M\"uhlenberg 1, 14476 Golm near Potsdam, 
Germany}
\date{{\small Preprint AEI-2000-027}} 

\begin{document}

\maketitle

\begin{abstract}
In this article we outline a rather general construction of  
diffeomorphism covariant coherent states for quantum gauge theories.

By this we mean states $\psi_{(A,E)}$, labelled by a point
$(A,E)$ in the classical phase space, consisting of canonically conjugate
pairs of connections $A$ and electric fields $E$ respectively, such 
that \\
(a) they are eigenstates of a corresponding annihilation operator which
is a generalization of $A-iE$ smeared in a suitable way,\\
(b) normal ordered polynomials of generalized annihilation and creation 
operators have the correct expectation value,\\ 
(c) they saturate the Heisenberg uncertainty bound for the fluctuations of
$\hat{A},\hat{E}$ and\\
(d) they do not use any background structure for their definition, that is,
they are diffeomorphism covariant.

This is the first paper in a series of articles entitled ``Gauge Field 
Theory Coherent States (GCS)'' which aim at connecting
non-perturbative quantum general relativity with the low energy physics of 
the standard model. In particular, coherent states enable us for the first
time to take into account quantum metrics which are excited {\it 
everywhere} in
an asymptotically flat spacetime manifold as is needed for semi-classical
considerations.

The formalism introduced in this paper
is immediately applicable also to lattice gauge theory in the presence of
a (Minkowski) background structure on a possibly {\it infinite lattice}.
\end{abstract}

\section{Introduction}
\label{s1}

Quantum General Relativity (QGR) has matured over the past decade to a 
mathematically well-defined theory of quantum gravity. 
In contrast to string theory, by definition QGR is a
manifestly background independent, diffeomorphism 
invariant and non-perturbative theory.
The obvious advantage is that one will never have to postulate the
existence of a non-perturbative extension of the theory,
which in string theory has been called the still unknown 
M(ystery)-Theory.

The disadvantage of a non-perturbative and background independent
formulation is, of course, that one is faced with new and interesting 
mathematical problems so that one cannot just go ahead and 
``start calculating scattering amplitudes'': 
As there is no background around which one could perturb, rather the full 
metric is fluctuating, one is not
doing quantum field theory on a spacetime but only on a differential
manifold. Once there is no (Minkowski) metric at our disposal, one loses
familiar notions such as causality, locality, Poincar\'e group 
and so forth, in other words, the theory is not a theory to which
the Wightman axioms apply. Therefore, one must build an entirely
new mathematical apparatus to treat the resulting quantum field theory 
which is {\it drastically different from the Fock space picture 
to which particle physicists are used to}.

As a consequence, the mathematical formulation of the theory was the main 
focus of research in the field over the past decade. The main 
achievements to date are the following (more or less in chronological 
order) : 
\begin{itemize}
\item[i)] {\it Kinematical Framework}\\
The starting 
point was the introduction of new field variables \cite{1} for the 
gravitational field which are better suited to a background  
independent formulation of the quantum theory than the ones employed
until that time. In its original version these variables were
complex valued, however, currently their real valued version,  
considered first in \cite{1a} for {\it classical} Euclidean gravity and 
later in \cite{1b} for {\it classical} Lorentzian gravity, is preferred 
because to date it seems that it is only with these variables that one can 
rigorously define the kinematics and dynamics of Euclidean or Lorentzian  
{\it quantum} gravity \cite{1c}. \\
These variables are coordinates for the infinite dimensional phase
space of an $SU(2)$ gauge theory subject to further constraints 
besides the Gauss law, that is, a connection and a canonically
conjugate electric field. As such, it is very natural to introduce
smeared functions of these variables, specifically Wilson loop and 
electric flux functions. (Notice that one does not need a metric 
to define these functions, that is, they are background independent).
This had been done for ordinary gauge fields already before in \cite{2} 
and was then reconsidered for gravity (see e.g. \cite{3}).\\
The next step was the choice of a representation of the canonical
commutation relations between the electric and magnetic degrees
of freedom. This involves the choice of a suitable space of 
distributional connections \cite{4} and a faithful measure thereon \cite{5}
which, as one can show \cite{6}, is $\sigma$-additive.
The proof that the resulting Hilbert space indeed solves the adjointness 
relations induced by the reality structure of the classical theory
as well as the canonical commutation relations induced by the symplectic 
structure of the classical theory can be found in \cite{7}.
Independently, a second representation of the canonical commutation
relations, called the loop representation, 
had been advocated (see e.g. \cite{8} and especially \cite{8a} and references 
therein)
but both representations were shown to be unitarily equivalent in
\cite{9} (see also \cite{10} for a different method of proof).\\
This is then the first major achievement : The theory is based on
a rigorously defined kinematical framework.
\item[ii)] {\it Geometrical Operators}\\
The second major achievement concerns the spectra of positive 
semi-definite, self-adjoint geometrical
operators measuring lengths \cite{11}, areas \cite{12,13}
and volumes \cite{12,14,15,16,8} of curves, surfaces and regions
in spacetime. These spectra are pure point (discete) and imply a discrete
Planck scale structure. It should be pointed out that the discreteness
is, in contrast to other approaches to quantum gravity, not put in
by hand but it is a {\it prediction} !
\item[iii)] {\it Regularization- and Renormalization Techniques}\\
The third major achievement is that there is a new 
regularization and renormalization technique \cite{17,18}
for diffeomorphism covariant, density-one-valued operators at our disposal
which was successfully tested in model theories \cite{19}. This
technique can be applied, in particular, to the standard model
coupled to gravity \cite{20,21} and to the Poincar\'e generators at 
spatial infinity \cite{22}. In particular, it works for {\it Lorentzian}
gravity while all earlier proposals could at best work in the Euclidean 
context only (see, e.g. \cite{8a} and references therein). 
The algebra of important operators of the
resulting quantum field theories was shown to be consistent \cite{23}. 
Most surprisingly, these operators are {\it UV and IR finite} !
Notice that this result, at least as far as these operators are 
concerned, is stronger 
than the believed but unproved finiteness of scattering amplitudes
order by order in perturbation theory of the five critical
string theories, in a sense we claim that the perturbation series converges.
The absence of the divergences that usually plague interacting quantum fields
propagating on a Minkowski background can be understood intuitively
from the diffeomorphism invariance of the theory : ``short and long distances
are gauge equivalent''. We will elaborate more on this point in future 
publications. 
\item[iv)] {\it Spin Foam Models}\\
After the construction of the densely defined Hamiltonian constraint
operator of \cite{17,18}, a formal, Euclidean functional integral was
constructed in \cite{23a} and gave rise to the so-called spin foam 
models   
(a spin foam is a history of a graph with faces as the history of edges)
\cite{23b}. Spin foam models are in close connection with causal
spin-network evolutions \cite{23c}, state sum models \cite{23d} and
topological quantum field theory, in particular BF theory \cite{23e}. To
date most results are at a formal level and for the Euclidean version of the
theory only but the programme is exciting since it may restore manifest
four-dimensional diffeomorphism invariance which in the Hamiltonian
formulation is somewhat hidden.
\item[v)]
Finally, the fifth major achievement is the existence of a rigorous and 
satisfactory framework \cite{24,25,26,27,28,29,30} for the quantum 
statistical description of black holes
which reproduces the Bekenstein-Hawking Entropy-Area relation and applies,
in particular, to physical Schwarzschild black holes while stringy black 
holes so far are under control only for extremal charged black holes.
\end{itemize}
Summarizing, the work of the past decade has now 
culminated in a promising starting point for a quantum theory of the 
gravitational field plus matter and the stage is set to pose and answer 
physical questions. 

The most basic and most important question that one should ask is :
{\it Does the theory have classical general relativity as its classical
limit ?} Notice that even if the answer is negative, the existence
of a consistent, interacting, diffeomorphism invariant quantum field theory 
in four dimensions is already a quite non-trivial result. However, we can 
claim to have a satisfactory quantum theory of Einstein's theory
only if the answer is positive. 

To settle this issue we have launched an attack based on coherent 
states which has culminated in a series of papers called 
``Gauge Field Theory Coherent States'' \cite{31,32,33,34,35,36}
and this paper is the first one in this collection which is going to be 
extended further.\\
\\
The organization of this series is the following :
\begin{itemize}
\item[I)] {\it General Properties}\\
In this paper we describe a fairly general method to generate 
families of diffeomorphism covariant coherent 
states with the usual desired properties such as annihilation
operator eigenstate nature, expectation value reproduction for
annihilation and creation operators and saturation of the Heisenberg 
uncertainty bound. If certain analytical conditions are met,
overcompleteness can be established as well.
The construction is based on the so-called configuration space 
complexifier method described in detail in \cite{36a}. The latter 
work arose as an abstraction
of the results of Hall \cite{36b} who chose a very special, but very 
convenient configuration space complexification for the case that the 
configuration space is a compact, connected Lie group. Hall's results
were later generalized to diffemorphism invariant gauge theories in 
\cite{36c}. In this paper we focus on general properties of such
states for a general complexification such as gauge invariance and 
diffeomorphism covariance. Besides such physical features also analytical 
properties are addressed and it is a mixture of the two that will
determine one's choice of the complexification. In fact, in the remainder
of this series we will mostly deal with a generalization of the 
complexification chosen by Hall. Our main reason for this  
choice is simply mathematical convenience : The spectrum of the 
operator that generates the configuration space complexification is 
explicitly known and sufficiently simple. This allows us to get started,
but it should be kept in 
mind that other choices are available that may prove physically more 
interesting later on in our programme.
\item[II)] {\it Peakedness Properties}\\
Associated with the configuration space complexification 
is a so-called coherent state transform and both of \cite{36b,36c}
focussed on the unitarity of that transform while the properties of  
the coherent states themselves remained untouched. Moreover, it remained 
unclear how the complexified connection $A^\Cl$ looks like in terms of the 
coordinates $(A,E)$ of the real phase space and without this an 
interpretation of the label $A^\Cl$ of the coherent state and thus 
expectation values, fluctuations and so forth remain veiled.
Here, $A$ is a connection for
a compact gauge group and $E$ is a canonically conjugate electric field.
To fill both of these gaps
is the purpose of the second paper \cite{31} in this series. 
First of all, we find the expected result, namely that roughly speaking
$A^\Cl=A-iE$ in a suitably smeared sense. Secondly,
we analyze in detail the peakedness properties of the coherent states for 
diffeomorphism invariant gauge theories in the configuration --, 
momentum --
and the Segal-Bargmann representation. We find that these states are 
very sharply peaked at the point $A$, $E$ or $(A,E)$ respectively
of the configuration --, momentum -- and phase space respectively.
That paper also contains extensive graphics to demonstrate these peakedness
properties pictorially and while there are important differences, the
states display the essential Gaussian decay of the harmonic oscillator
coherent states. 
\item[III)] {\it Ehrenfest Theorems}\\
In the third paper \cite{32} of this series we prove Ehrenfest theorems
for our coherent states. That is, we show that the expectation value
not only of normal ordered polynomials of creation an annihilation operators 
but of all polynomials of the elementary operators associated with
$\hat{A},\hat{E}$ equals, to leading order in $\hbar$, precisely the labels
$A,E$ of the coherent state. This result can be extended to certain 
operators that are non-polynomial in the basic ones and that appear
in the Hamiltonian constraint of quantum general relativity coupled to 
matter \cite{17,20,21}. Moreover, we show that commutators between
these operators divided by $i\hbar$ have an expectation value which 
equals to leading order in $\hbar$ the correspending Poisson bracket 
evaluated at the label $(A,E)$ of the coherent state. Together, these 
results imply that the classical limit of the Hamiltonian constraint 
operator and its infinitesimal quantum dynamics correspond to its
classical counterparts. 

Both of \cite{31,32} mainly deal with $G=U(1),SU(2)$ but we sketch  
how all the results can be extended to groups of higher rank, an issue
which we will examine in detail in \cite{36d}.
\item[IV)] {\it Infinite Tensor Product and Thermodynamical Limit}\\
The states that one considered in Quantum General Relativity until now
are labelled 
by piecewise analytic, {\it finite} graphs (an extension to {\it finite} 
collections of smooth curves with controlled intersection properties is 
possible, see later on). However, finite graphs are suitable to describe 
semiclassical physics on physically interesting spacetimes 
only if the underlying manifold is spatially compact. The most interesting
applications, flat space or an entire black hole spacetime (and not 
only the horizon region) cannot be treated
with finite graphs. To extend the framework it turns out that piecewise
analytical, countably infinite graphs together with the framework of
the Infinite Tensor Product (ITP) construction introduced by von Neumann 
\cite{36e} more than sixty years ago are appropriate. 
To the best of the knowledge of the author, the first time that 
truly infinite graphs and infinite tensor product states were considered in 
QGR in the context of a Hilbert space structure, was in section 3.2 of 
\cite{22} which dealt with the asymptotic Poincar\'e group of 
asymptotically flat spacetimes, however, the overall mathematical framework
of such constructions was not described there.
In \cite{33} we deliver this structure and embed it into our coherent 
states
framework. In particular, we are able to connect mathematical notions with
physical ones, an example being the following :\\ 
A state $f$ in the infinite tensor product Hilbert space over an 
infinite graph which is a direct product of normalized states, one for each
edge of the graph, generates so-called strong and weak equivalence classes 
of so-called $C_0$-sequences. It turns out that the corresponding 
$C_0$-vector plays the role of 
a cyclic vector (vacuum state) for a Fock-like tiny closed subspace of the 
complete ITP Hilbert space, called an $f$-adic incomplete ITP. Fock-like 
spaces corresponding to different strong and weak
equivalence classes are mutually orthogonal. Those Fock-like spaces that
correspond to the same weak class but different strong classes are 
unitarily equivalent while those that correspond to different strong and 
weak classes are unitarily inequivalent.  This way the ITP
gives rise to an uncountably infinite number of mutually unitarily 
inequivalent representations of the canonical commutation relations.
The representation theory of operator algebras becomes especially 
interesting, the enveloping framework being that of factors of 
von Neumann algebras.

Generically, incomplete ITP's generated by different weak equivalence 
classes correspond to physical situations which differ drastically
with respect to certain physical quantities such as energy, volume or  
topology. For instance, the Ashtekar-Isham-Lewandowski Hilbert space
based on finite graphs describes finite volume and/or compact topology
while a $C_0$ vector of infinite volume can be constructed by using our 
coherent states, appropriate to approximate a flat Minkowski space geometry.
The two Hilbert spaces are mutually orthogonal closed subspaces within
our complete ITP Hilbert space corresponding to different weak classes. The 
vacuum underlying the 
Ashtekar-Isham-Lewandowski Hilbert space via the GNS construction is based 
on a $C_0$ vector which equals unity 
for every edge of any possible graph. It can be shown that such a state,
in the context of non-compact topologies, is a pure quantum 
vacuum in the sense that it describes metrics of almost
everywhere zero spatial volume. 

It should be clear from 
these considerations that the ITP is possibly able to {\it describe all 
phyically different situations
at once and might enable us to describe topology change within
canonical quantum general relativity and therefore to get rid off the 
embedding spacetime manifold that one started with classically} !

The infinite tensor product opens the gate to a plethora of other physical
and mathematical disciplines, such as thermodynamics and statistical 
field theory, Tomita-Takesaki (or modular) theory necessary to classify  
the appearing types of type III factors of von Neumann algebras etc. 
\item[V)] {\it Higgs Fields and Fermions}\\
The framework described so far is sufficient for pure quantum gauge theories
coupled to quantum general relativity only. By combining the framework of
\cite{21} with the infinite tensor product construction and existing
results for coherent states for fermions (e.g. \cite{36f} and references 
therein) we can extend the framework to all matter of the standard model
including possible supersymmetric extensions. The details are described
in \cite{34}.
\item[VI)] {\it Photons and Gravitons}\\
Most of the criticism directed towards quantum general relativity coming from 
the particle physics community is that the programme, being manifestly 
non-perturbative by construction, seems to be infinitely far away from 
any established perturbative results such as (free) quantum field theory
on curved backgrounds (widely believed to be the first approximation
to full quantum gravity), perturbative quantum (super)gravity 
(non-renormalizable) and perturbative quantum superstring theory.
In \cite{35} we make a first contact with these programmes. Namely,
we try to construct a map between the perturbative Photon or Graviton 
Hilbert 
spaces and a fully non-perturbative incomplete $f$-adic ITP subspace where
the $C_0$-vector corresponding to $f$ is a best approximation state
to the Minkowski space solution of the Einstein-Maxwell equations. This work 
is aimed at demonstrating how perturbative notions such as particles can be 
absorbed into our fully non-perturbative programme.
\item[VII)] {\it The Non-Perturbative $\gamma$-Ray Burst Effect}\\
Many serious theorists and experimentalists nowadays discuss the possibility 
to actually measure quantum gravity effects, a prominent example being the 
so-called $\gamma$-ray burst experiment (see, e.g. \cite{36g,36h}).
In all these types of experiments one exploits the fact that the incredibly
tiny quantum gravity effects may accumulate over vast periods of time
of the order of the age of the universe to a measurable size. 
In particular, the theoretical mechanism of the $\gamma$-ray burst effect 
can be roughly described as follows : the quantum metric depends on 
canonically conjugate magnetic and electric degrees of freedom and thus the 
Heisenberg uncertainty obstruction tells us that there is no state
that can describe the Minkowski vacuum {\it exactly}. In other words,
there is no Poincar\'e invariant state in the theory, the best one
can do is to construct a coherent state peaked on Minkowski space.
The expectation value of the Einstein-Maxwell-Hamiltonian with respect
to the gravitational field will therefore include corrections to the 
classical Minkowski metric which give rise to Poincar\'e invariance
violating dispersion relations. Thus, if one could measure the 
arrival times of $\gamma$-ray photons of different energies they should   
differ by an amount proportional to the travelling time from the source.

The challenge is now to precisely compute
these corrections from our fully non-perturbative framework, in particular,
what is the precise power of the Planck mass that the effect is
proportional to. This is the subject of \cite{36} which will improve the 
pioneering work \cite{36k} in two respects : First, the latter was based on 
so-called weave states \cite{36l} which,
however, approximate only half of the number of degrees of freedom and,
secondly, in contrast to our coherent states the existence of weave itself 
with the assumed semi-classical properties was not proved to exist.

To compute the effect exactly turns out to be a hard piece of analysis due to 
the non-linear, even non-analytic 
(interacting) nature of the theory, a property which carries over to 
our coherent states. In particular, the complicated spectrum of the volume 
operator makes the enterprise not an easy one. On the other hand,
it is absolutely crucial to know the {\it precise} spectrum and not only of, 
say, its main series : If one would do the same with the area operator
then, as has been beautifully demonstrated in \cite{30}, one would 
reach the conclusion that the black hole Hawking radiation spectrum is
discrete rather than the quasi-continuous one of a black body, in other words,
the spectrum has direct bearing on observation !

It is at this point that super-computers may
enter the stage as analytic computations start becoming 
too hard and lengthy. Notice, however, that in contrast to usual 
perturbation series in perturbative quantum field theory the computational 
error is always under good control. The series that we are dealing with are 
manifestly absolutely converging and there are precise estimates on
the error that one creates when keeping only the dominant terms. We will
display such error controlled estimates in the next two issues of this 
series. 
\item[VIII)] {\it The Classical Limit}\\
As an immediate application of coherent states and the ITP framework
one can now precisely prove in detail \cite{36m} how it happens 
that the Hamiltonian constraint constructed in \cite{17} obeys the correct
quantum algebra.
\end{itemize}
More work is in progress. The following list of projects 
associated with our coherent states represents just the tip of the iceberg,
in principle it would would be interesting to repeat all 
perturbative calculations that
have been performed so far with our non-perturbative tools and to provide 
the error bars. \\
A)\\
To relate standard perturbative quantum field theory on curved backgrounds 
with non-perturbative quantum general relativity one would like to 
understand why the UV singularities of the former have disappeared 
in the latter. The naive answer is that the renormalization group 
has been absorbed into the diffeomorphism group (large and small momenta
are gauge related) but one would like to understand this and related
notions like bare and renormalized charges, effective actions,
renormalization transformations, Epstein-Glaser formalism and the importance 
of Hadamard states for quantum field theory on curved backgrounds etc. in 
more detail from the non-perturbative point of view. In particular, it would 
be nice to map the usual Feynman rules into our framework. This
research project will be started in \cite{36n,36o}.\\
B)\\
An ever fascinating research object has been the black hole. The coherent
states provide a natural new setting in which to study quantum black holes 
and Hawking radiation, in principle one ``just'' has to take the coherent
state that approximates a Kruskal spacetime together with its excitations
in order to provide the Kruskal -- spacetime -- adic incomplete closed ITP 
Hilbert space structure (that is, a vacuum and excitations). Notice that 
while the Bekenstein -- Hawking entropy 
has been successfully computed in both canonical quantum gravity and string
theory as mentioned above, what would be new here is that one can treat 
the full spacetime in a Hilbert space context and not only its near horizon 
structure (charges). Also, there are a priori no constraints such as
(near-) stationarity or extremality of the black hole.
Finally, one would like to understand what happens to the classical 
singularity theorems, the information paradoxon, cosmic censorship etc. in 
the quantum theory.
These and rlated issues will be the topic of \cite{36p}.\\
C)\\
As already mentioned, von Neumann algebras and their representation theory
appear quite naturally in the Infinite Tensor Product construction.
For the latter, the decomposition of a von Neumann algebra into factors
is of particular importance and the basic tool to characterize 
factors of type III, which typically appear in quantum field theory, is
provided by modular theory. This brings us into close 
contact with algebraic quantum field theory, although presumably in a 
generalized setting, since the notion of locality plays, almost by 
definition, a less dominant role in a diffeomorphism invariant quantum 
field theory. These and related issues will be examined in \cite{36q}. \\
D)\\ 
The most effective way to derive a path integral formulation for 
kinematically linear field theories from the 
Hamiltonian formulation of the theory is via coherent states, see e.g.
\cite{36f} and references therein. Thus, it is natural to expect this
to be the case also for our coherent states. This may bring us into
contact with the formely mentioned spin foam models 
\cite{23a,23b,23c,23d,23e} which have
recently attracted quite some attention after the appearence of \cite{17,18}
and will be studied in \cite{36r}.\\
E)\\
Finally, our coherent states are pure states. The semi-classical 
behaviour of such states may yet be improved by superimposing them to a 
so-called mixed state which makes use of random lattices. For weaves,
such a framework already exists and has been studied in \cite{36s}.
We intend to combine both frameworks in \cite{36t}.\\
\\
\\
This article is assembled as follows :\\
\\

In section two we recall the classical and quantum kinematics of 
diffeomorphism invariant gauge field theories.

In section three we recall the complexifier method to generate 
Bargmann-Segal representations for general theories and gauge theories 
in particular. We comment on the physical and mathematical requirements to 
be imposed on the complexifier, that is, the canonical generator of the 
transform that complexifies the configuration space and identifies it with
the phase space. In three related subsections we propose three candidate
families of coherent states for gauge theories. The first one leads to
an actual complex connection, the second only to a complexified holonomy 
without underlying complex connection and the third one maps the problem
at hand in principle to coherent states for an (in)finite collection of 
uncoupled harmonic oscillators. We describe the advantages and disadvantages
of these states as compared to each other. All of this will be done mostly 
for gauge -- and diffeomorphism variant coherent states.

In sections four and five respectively we will deal with the issue of 
how to construct gauge -- and diffeomorphism invariant coherent states 
respectively. Some of these can even be chosen to be annihilated by 
the Hamiltonian constraint.

Finally, in section six we display a simple example for gauge invariant
coherent states with an actual complex connection in 2+1 gravity and study
some of their peakedness properties.

\section{Kinematical Structure of Diffeomorphism Invariant Quantum
Gauge Theories}
\label{s2}

In this section we will recall the main ingredients of the mathematical
formulation of (Lorentzian) diffeomorphism invariant classical and quantum 
field theories of 
connections with local degrees of freedom in any dimension and for
any compact gauge group. See \cite{7,37} and references therein
for more details.

\subsection{Classical Theory}
\label{s2.1}

Let $G$ be a compact gauge group, $\Sigma$ a $D-$dimensional manifold 
admitting a principal $G-$bundle with connection over $\Sigma$.
Let us denote the pull-back to $\Sigma$ of the connection 
by local sections by $A_a^i$
where $a,b,c,..=1,..,D$ denote tensorial indices and $i,j,k,..=1,..,
\dim(G)$ denote indices for the Lie algebra of $G$. 
Likewise, consider a density-one vector bundle of electric fields, whose
pull-back to $\Sigma$ by local sections (their Hodge dual is a 
$D-1$ form) is a Lie algebra valued 
vector density of weight one. We will denote the set of generators
of the rank $N-1$ Lie algebra of $G$ by $\tau_i$ which are normalized
according to $\mbox{tr}(\tau_i\tau_j)=-N\delta_{ij}$ and 
$[\tau_i,\tau_j]=2f_{ij}\;^k\tau_k$ defines the structure constants 
of $Lie(G)$. 

Let $F^a_i$ be a Lie algebra valued vector density test field of weight one 
and let $f_a^i$ be a Lie algebra valued covector test field. 
We consider the smeared quantities
\be \label{2.1}
F(A):=\int_\Sigma d^Dx F^a_i A_a^i\mbox{ and } 
E(f):=\int_\Sigma d^Dx E^a_i f_a^i 
\ee
While both of them are diffeomorphism covariant it is only the latter 
which is gauge covariant and this is one motivation to consider the 
singular smearings discussed below.
The choice of the space of pairs of test fields $(F,f)\in{\cal S}$ 
depends on the boundary conditions on
the space of connections and electric fields which in turn depends on the 
topology of $\Sigma$ and will not be specified in what follows. 

Let the set
of all pairs of smooth functions $(A,E)$ on $\Sigma$ such that (\ref{2.1}) is 
well defined for any $(F,f)\in {\cal S}$ be denoted by
$M$. We define a topology on $M$ through the following globally 
defined metric : 
\ba \label{2.2}
&& d_{\rho,\sigma}[(A,E),(A',E')] \\
&:=& \sqrt{-\frac{1}{N}\int_\Sigma d^Dx 
[\sqrt{\det(\rho)} \rho^{ab} \mbox{tr}([A_a-A'_a][A_b-A'_b])+
\frac{[\sigma_{ab} \mbox{tr}([E^a-E^{a\prime}][E^b-E^{b\prime}])}
{\sqrt{\det(\sigma)}}]} \nonumber
\ea
where $\rho_{ab},\sigma_{ab}$ are fiducial metrics on $\Sigma$ of 
everywhere Euclidean signature. Their fall-off behaviour has to be suited
to the boundary conditions of the fields $A,E$ at spatial infinity.
Notice that the metric (\ref{2.2}) on $M$ is gauge invariant. It can be used   
in the usual way to equip $M$ with the structure of a smooth,
infinite dimensional differential
manifold modelled on a Banach (in fact Hilbert) space $\cal E$
where ${\cal S}\times {\cal S}\subset {\cal E}$. (It is the 
weighted Sobolev space $H_{0,\rho}^2\times H_{0,\sigma^{-1}}^2$ in the 
notation of \cite{38}). 

Finally, we equip $M$ with the structure of an infinite dimensional 
symplectic manifold through the following strong (in the sense of 
\cite{39})
symplectic structure 
\be \label{2.3}
\Omega((f,F),(f',F'))_m:=\int_\Sigma d^Dx [F^a_i f^{i\prime}_a
-F^{a\prime}_i f^i_a](x)
\ee
for any $(f,F),(f',F')\in {\cal E}$. We have abused the notation by 
identifying the tangent space to $M$ at $m$ with $\cal E$. To prove 
that $\Omega$ is a strong symplectic structure one uses standard 
Banach space techniques. Computing the Hamiltonian vector fields
(with respect to $\Omega$) of the functions $E(f),F(A)$ we obtain the
following elementary Poisson brackets
\be \label{2.4}
\{E(f),E(f')\}=\{F(A),F'(A)\}=0,\;\{E(f),A(F)\}=F(f)
\ee
As a first step towards quantization of the symplectic manifold
$(M,\Omega)$ one must choose a polarization. As usual in gauge theories,
we will use a particular real polarization, specifically connections as 
the configuration variables and electric fields 
as canonically conjugate momenta. As a second step one must decide
on a complete set of coordinates of $M$ which are to become the elementary
quantum operators. The analysis just outlined suggests to use the 
coordinates $E(f),F(A)$. However, the well-known immediate problem is that 
these coordinates are not gauge covariant. Thus, we proceed as follows :

Let $\Gamma^\omega_0$ be the set
of all piecewise analytic, finite, oriented graphs $\gamma$ embedded into 
$\Sigma$ 
and denote by $E(\gamma)$ and $V(\gamma)$ respectively its sets of oriented
edges $e$ and vertices $v$ respectively. Here finite means that 
$E(\gamma)$ is a finite set. (One can extend the framework to 
$\Gamma^\infty_0$, the restriction to webs of the set of
piecewise smooth graphs \cite{40,41} but the description becomes more 
complicated and we refrain from doing this here). 
It is possible to consider the set $\Gamma^\omega_\sigma$ of piecewise 
analytic, infinite graphs with
an additional regularity property \cite{34} but for the purpose of this
paper it will be sufficient to stick to $\Gamma^\omega_0$. The subscript
$_0$ as usual denotes ``of compact support'' while $_\sigma$ denotes
``$\sigma$-finite''.

We denote by $h_e(A)$ the holonomy
of $A$ along $e$ and say that a function $f$ on $\a$ is cylindrical with 
respect to $\gamma$ if there exists a function $f_\gamma$ on 
$G^{|E(\gamma)|}$ such that $f=p_\gamma^\ast f_\gamma
=f_\gamma\circ p_\gamma$ 
where $p_\gamma(A)=\{h_e(A)\}_{e\in E(\gamma)}$. 
Holonomies are invariant under
reparameterizations of the edge and in this article we assume that
the edges are always analyticity preserving diffeomorphic images from 
$[0,1]$ to a
one-dimensional submanifold of $\Sigma$. Gauge transformations are functions
$g:\;\Sigma\mapsto G;\;x\mapsto g(x)$ and they act on
holonomies as $h_e\mapsto g(e(0))h_e g(e(1))^{-1}$. 

Next, given a graph $\gamma$ we choose a polyhedronal decomposition
$P_\gamma$ of $\Sigma$ dual to $\gamma$. The precise definition
of a dual polyhedronal decomposition can be found in \cite{37} but
for the purposes of the present paper it is sufficient to know that
$P_\gamma$ assigns to each edge $e$ of $\gamma$ an open ``face''
$S_e$ (a polyhedron of codimension one embedded into $\Sigma$) with 
the following properties :\\ 
(1) the surfaces $S_e$ are mutually non-intersecting,\\ 
(2) only the edge $e$ intersects $S_e$, the intersection is transversal
and consists only of one point which is an interior point of both
$e$ and $S_e$,\\
(3) $S_e$ carries the orientation which agrees with the orientation 
of $e$.\\
Furthermore, we choose a system $\Pi_\gamma$ of paths $\rho_e(x) \subset
S_e,\; x\in S_e,\; e\in E(\gamma)$ connecting the intersection point 
$p_e=e\cap S_e$ with $x$. The paths vary smoothly with
$x$ and the triples $\gamma,P_\gamma,\Pi_\gamma$
are such that if $\gamma,\gamma'$ are diffeomorphic, so
are $P_\gamma,P_{\gamma'}$ and $\Pi_\gamma,\Pi_{\gamma'}$, see
\cite{37} for details.

With these structures we define the following function on $(M,\Omega)$
\be \label{2.5}
P^e_i(A,E):=-\frac{1}{N}
\mbox{tr}(\tau_i h_e(0,1/2)[\int_{S_e} h_{\rho_e(x)} \ast E(x) 
h_{\rho_e(x)}^{-1}] h_e(0,1/2)^{-1})
\ee
where $h_e(s,t)$ denotes the holonomy of $A$ along $e$ between the 
parameter values $s<t$, $\ast$ denotes the Hodge dual, that is,
$\ast E$ is a $(D-1)-$form on $\Sigma$, $E^a:=E^a_i\tau_i$ and
we have chosen a parameterization of $e$ such that $p_e=e(1/2)$.

Notice that in contrast to similar variables used earlier in the literature
the function $P^e_i$ is {\it gauge covariant}. Namely, under gauge 
transformations it transforms as $P^e\mapsto g(e(0)) P^e g(e(0))^{-1}$,
the price to pay being that $P^e$ depends on both $A$ and $E$ and not 
only on $E$. The idea is therefore to use the variables $h_e,P^e_i$
for all possible graphs $\gamma$ as the coordinates of $M$.

The problem with the functions $h_e(A)$ and $P^e_i(A,E)$ on $M$ is that 
they are not differentiable on $M$, that is, $Dh_e, DP^e_i$ are nowhere  
bounded operators on $\cal E$ as one can easily see. The reason for this is,
of course, that these are functions on $M$ which are not properly smeared 
with functions from $\cal S$, rather they are smeared with distributional
test functions with support on $e$ or $S_e$ respectively. Nevertheless
one would like to base the quantization of the theory on these functions 
as basic variables because of their gauge and diffeomorphism covariance.
Indeed, under diffeomorphisms $h_e\mapsto h_{\varphi^{-1}(e)},
P^e_i\mapsto P^{\varphi^{-1}(e)}_i$ where the latter notation means that
$P^{\varphi^{-1}(e)}_e$ is labelled by 
$\varphi^{-1}(S_e),\varphi^{-1}(\Pi_\gamma)$. We proceed as follows. 
\begin{Definition} \label{def2.1}
By $\bar{M}_\gamma$ we denote the direct product 
$[G\times Lie(G)]^{|E(\gamma)|}$. 
The subset of $\bar{M}_\gamma$ of pairs $(h_e(A),P^e_i(A,E))_{e\in 
E(\gamma)}$ as 
$(A,E)$ varies over $M$ will be denoted by $(\bar{M}_\gamma)_{|M}$. We 
have a corresponding map $p_\gamma:\;M\mapsto \bar{M}_\gamma$ which
maps $M$ onto $(\bar{M}_{\gamma})_{|M}$.
\end{Definition}
Notice that the set $(\bar{M}_\gamma)_{|M}$ is in general a proper subset of 
$\bar{M}_\gamma$,
depending on the boundary conditions on $(A,E)$, the topology of $\Sigma$ 
and the ``size'' of $e,S_e$. For instance, in the limit of $e,S_e\to  
e\cap S_e$ but holding the number of edges fixed, $(\bar{M}_\gamma)_{|M}$  
will consist of only one point in $M_\gamma$. This follows from the 
smoothness of the $(A,E)$. 

We equip a subset $M_\gamma$ of $\bar{M}_\gamma$ with the structure of a 
differentiable manifold 
modelled on the Banach space ${\cal E}_\gamma=\Rl^{2\dim(G)|E(\gamma)|}$
by using the natural direct product manifold structure of
$[G\times Lie(G)]^{|E(\gamma)|}$. While $\bar{M}_\gamma$ is a kind of 
distributional phase space, $M_\gamma$ satisfies appropriate regularity
properties induced by (\ref{2.2}). 

In order to proceed and to give $M_\gamma$ a symplectic structure 
{\it derived from $(M,\Omega)$} one must 
regularize the elementary functions $h_e, P^e_i$ by writing them as limits  
(in which the regulator vanishes) of functions which can be expressed  
in terms of the $F(A),E(f)$. Then one can compute their Poisson
brackets with respect to the symplectic structure $\Omega$ at finite
regulator and then take the limit pointwise on $M$. The result is the 
following  
well-defined strong symplectic structure $\Omega_\gamma$ on $M_\gamma$. 
\ba \label{2.6}
\{h_e,h_{e'}\}_\gamma &=& 0\nonumber\\
\{P^e_i,h_{e'}\}_\gamma &=&
\delta^e_{e'} \frac{\tau_i}{2}h_e\nonumber\\
\{P^e_i,P^{e'}_j\}_\gamma &=&
-\delta^{ee'}f_{ij}\;^k P^e_k
\ea
Since $\Omega_\gamma$ is obviously block diagonal, each block standing
for one copy of $G\times Lie(G)$, to check that $\Omega_\gamma$ is 
non-degenerate and closed reduces to doing it for each factor together
with an appeal to well-known Hilbert space techniques to establish that
$\Omega_\gamma$ is a surjection of ${\cal E}_\gamma$.
This is done in \cite{37} where it is shown that each copy is isomorphic
with the cotangent bundle $T^\ast G$ equipped with the symplectic structure
(\ref{2.6}) (choose $e=e'$ and delete the label $e$). \\
\\
Now that we have managed to assign to each graph $\gamma$ a symplectic
manifold $(M_\gamma,\Omega_\gamma)$ we can quantize it by using geometric
quantization. This can be done in a well-defined way because the relations
(\ref{2.6}) show that the corresponding operators are non-distributional.
This is therefore a clean starting point for the regularization of any 
operator
of quantum gauge field theory which can always be written in terms 
of the $\hat{h}_e,\hat{P}^e,\;e\in E(\gamma)$ if we apply this operator to
a function which depends only on the $h_e,\; e\in E(\gamma)$. 

The question is what $(M_\gamma,\Omega_\gamma)$ has to do with $(M,\Omega)$.
In \cite{37} it is shown that there exists a partial order $\prec$ on the 
set $\cal L$ of triples $l=(\gamma,P_\gamma,\Pi_\gamma)$. 
In particular, $\gamma\prec\gamma'$ means $\gamma\subset\gamma'$
and $\cal L$ is a directed set so that one can form
a generalized projective limit $M_\infty$ of the $M_\gamma$ (we abuse 
notation in 
displaying the dependence of $M_\gamma$ on $\gamma$ only rather than on
$l$). For this one verifies that the family 
of symplectic structures $\Omega_\gamma$ is self-consistent
in the sense that if 
$(\gamma,P_\gamma,\Pi_\gamma)\prec (\gamma',P_{\gamma'},\Pi_{\gamma'})$ 
then $p_{\gamma'\gamma}^\ast\{f,g\}_\gamma
=\{p_{\gamma'\gamma}^\ast f,p_{\gamma'\gamma}^\ast g\}_{\gamma'}$
for any $f,g\in C^\infty(M_\gamma)$ and 
$p_{\gamma'\gamma}:\;M_{\gamma'}\mapsto M_\gamma$ is a system
of natural projections, more precisely, of (non-invertible) 
symplectomorphisms. 

Now, via the maps $p_\gamma$ of definition \ref{def2.1} we can identify
$M$ with a subset of $M_\infty$. Moreover, in \cite{37} it is shown that
there is a generalized projective sequence $(\gamma_n,P_{\gamma_n},
\Pi_{\gamma_n})$
such that $\lim_{n\to\infty}p_{\gamma_n}^\ast\Omega_{\gamma_n}=\Omega$
pointwise in $M$. This displays $(M,\Omega)$ as embedded into a
generalized projective
limit of the $(M_\gamma,\Omega_\gamma)$, intuitively speaking, as $\gamma$ 
fills all of $\Sigma$, we recover $(M,\Omega)$ from the 
$(M_\gamma,\Omega_\gamma)$. Of course, this works with $\Gamma^\omega_0$ 
only if $\Sigma$ is compact, otherwise we need the extension to 
$\Gamma^\omega_\sigma$.

It follows that quantization of $(M,\Omega)$, and conversely taking the 
classical limit, can be studied purely in terms of $(M_\gamma,\Omega_\gamma)$
for {\it all} $\gamma$. The quantum kinematical framework is
given in the next subsection.

\subsection{Quantum Theory}
\label{s2.2}

Let us denote the set of all smooth connections by $\a$. This is our
classical configuration space and we will choose for its coordinates the
holonomies $h_e(A),\;e\in\gamma,\;\gamma\in\Gamma^\omega_0$. 
$\a$ is naturally equipped with a metric topology induced by (\ref{2.2}). 

Recall the notion of a function cylindrical over a graph from the 
previous subsection.
A particularly useful set of cylindrical functions are the so-called 
spin-netwok functions \cite{42,43,9}. A spin-network function is 
labelled by a graph $\gamma$, a set of non-trivial irreducible 
representations 
$\vec{\pi}=\{\pi_e\}_{e\in E(\gamma)}$ (choose from each equivalence 
class of equivalent
representations once and for all a fixed representant), one for each 
edge of $\gamma$, and a set $\vec{c}=\{c_v\}_{v\in V(\gamma)}$ of
contraction matrices, one for each vertex of $\gamma$, which 
contract the indices of the tensor product 
$\otimes_{e\in E(\gamma)} \pi_e(h_e)$ in such a way that the resulting
function is gauge invariant. We denote spin-network functions as
$T_I$ where $I=\{\gamma,\vec{\pi},\vec{c}\}$ is a compound label.
One can show that these functions are linearly independent.
From now on we denote by $\tilde{\Phi}_\gamma$ finite linear combinations of
spin-network functions over $\gamma$, by $\Phi_\gamma$ the finite linear 
combinations of elements from any possible $\tilde{\Phi}_{\gamma'},\;
\gamma'\subset\gamma$ a subgraph of $\gamma$  
and by $\Phi$ the finite linear 
combinations of spin-network functions over an arbitrary finite collection 
of graphs. Clearly $\tilde{\Phi}_{\gamma}$ is a subspace of 
$\Phi_\gamma$.  To express this distinction we will say that functions 
in $\tilde{\Phi}_\gamma$ are labelled by the ``coloured graphs'' $\gamma$
while functions in $\Phi_\gamma$ are labelled simply by graphs $\gamma$
where we abuse notation by using the same symbol $\gamma$.

The set $\Phi$ of finite linear combinations of spin-network functions 
forms an Abelian $^\ast$ algebra 
of functions on $\a$. By completing it with respect to the sup-norm 
topology it 
becomes an Abelian C$^\ast$ algebra $\cal B$ (here the compactness of $G$ is 
crucial). The spectrum $\ab$ of this algebra, 
that is, the set of all algebraic homomorphisms ${\cal B}\mapsto\Cl$
is called the quantum configuration space. This space is equipped with
the Gel'fand topology, that is, the space of continuous functions
$C^0(\ab)$
on $\ab$ is given by the Gel'fand transforms of elements of $\cal B$.
Recall that the Gel'fand transform is given by $\tilde{f}(\bar{A}):=
\bar{A}(f)\;\forall \bar{A}\in \ab$. It is a general result that $\ab$ with 
this topology is a compact Hausdorff space. Obviously, the elements of
$\a$ are contained in $\ab$ and one can show that $\a$ is even dense
\cite{44}. Generic elements of $\ab$ are, however, distributional.

The idea is now to construct a Hilbert space consisting of square
integrable functions on $\ab$ with respect to some measure $\mu$. Recall 
that one can define a measure on a locally compact Hausdorff space 
by prescribing a positive linear functional $\chi_\mu$ on the space 
of continuous functions thereon. The particular measure
we choose is given by $\chi_{\mu_0}(\tilde{T}_I)=1$ if $I=\{\{p\},
\vec{0},\vec{1}\}$ and $\chi_{\mu_0}(\tilde{T}_I)=0$ otherwise. Here
$p$ is any point in $\Sigma$, $0$ denotes the 
trivial representation and $1$ the trivial contraction matrix. In other 
words, (Gel'fand transforms of) spin-network functions play the same role 
for $\mu_0$ as 
Wick-polynomials do for Gaussian measures and like those they form
an orthonormal basis in the Hilbert space ${\cal H}:=L_2(\ab,d\mu_0)$ 
obtained by completing their finite linear span $\Phi$.\\
An equivalent definition of $\ab,\mu_0$ is as follows :\\ 
$\ab$ is in one to one correspondence, via the surjective map $H$ defined 
below, with the set $\ab':=\mbox{Hom}({\cal X},G)$
of homomorphisms from the groupoid $\cal X$ of composable, holonomically
independent, analytical paths
into the gauge group. The correspondence is explicitly given by
$\ab\ni\bar{A}\mapsto H_{\bar{A}}\in\mbox{Hom}({\cal X},G)$
where ${\cal X}\ni e\mapsto H_{\bar{A}}(e):=\bar{A}(h_e)=
\tilde{h}_e(\bar{A})\in G$ and $\tilde{h}_e$ is the Gel'fand transform
of the function $\a\ni A\mapsto h_e(A)\in G$. Consider now the restriction
of $\cal X$ to ${\cal X}_\gamma$, the groupoid of composable edges of  
the graph $\gamma$. One can then show that the projective limit 
of the corresponding {\it cylindrical sets} 
$\ab'_\gamma:=\mbox{Hom}({\cal X}_\gamma,G)$ coincides with $\ab'$.
Moreover, we have $\{\{H(e)\}_{e\in E(\gamma)};\;H\in\ab'_\gamma\}=
\{\{H_{\bar{A}}(e)\}_{e\in E(\gamma)};\;\bar{A}\in\ab\}=
G^{|E(\gamma)|}$.
Let now $f\in{\cal B}$ be a function cylindrical over $\gamma$ then 
$$
\chi_{\mu_0}(\tilde{f})=\int_{\ab} d\mu_0(\bar{A}) \tilde{f}(\bar{A})
=\int_{G^{|E(\gamma)|}} \otimes_{e\in E(\gamma)} d\mu_H(h_e)
f_\gamma(\{h_e\}_{e\in E(\gamma)})
$$
where $\mu_H$ is the Haar measure on $G$.
As usual, $\a$ turns out to be contained in a measurable subset of 
$\ab$ which has measure zero with respect to $\mu_0$.

Let $\Phi_\gamma$, as before, be the finite linear span of spin-network 
functions
over $\gamma$ and ${\cal H}_\gamma$ its completion with respect to
$\mu_0$. Clearly, $\cal H$ itself is the completion of the finite linear
span $\Phi$ of vectors from the mutually orthogonal $\tilde{\Phi}_\gamma$. 
Our 
basic coordinates of $M_\gamma$ are promoted to operators on ${\cal H}$ with 
dense domain $\Phi$. As $h_e$ is group-valued and $P^e$ is real-valued
we must check that the adjointness relations coming from these reality 
conditions as well as the Poisson brackets (\ref{2.6}) are implemented on
our ${\cal H}$. This turns out to be precisely the case if we choose
$\hat{h}_e$ to be a multiplication operator and 
$\hat{P}^e_j=i\hbar\kappa X^e_j/2$
where $X^e_j=X_j(h_e)$ and $X^j(h),\;h\in G$ is the vector field on $G$
generating left translations into the $j-th$ coordinate direction of 
$Lie(G)\equiv T_h(G)$ (the tangent space of $G$ at $h$ can be identified 
with the Lie algebra of $G$) and $\kappa$ is the coupling constant of the 
theory. For details see \cite{7,37}.

\section{Coherent States from a Coherent State Transform}
\label{s3}

In the first subsection of this section we will recall the state of the art
of families of coherent state transforms which have been defined in the 
literature already. We point out advantages and disadvantages  
of one transform as compared to another as well as general properties of
every transform and draw attention to some gaps that 
were left over. In the subsequent subsection we show how some of these gaps
can be filled.

\subsection{Review of Known Results}
\label{s3.1}

The first construction of coherent states that are relevant for the
quantization of cotangent bundles over connected compact Lie groups $G$
is due to Hall \cite{36b} who showed how to construct a unitary
map between the Hilbert space $L_2(G,d\mu_H)$ and a Hilbert space
consisting of square integrable holomorphic functions of the
complexification $G^\Cl$ of $G$ with respect to some measure $\nu$ that
he explicitly constructed. In \cite{36c} these results were applied
to our graph theoretic framework, namely one needs to repeat Hall's
construction, roughly speaking, for every holonomy associated with the
various edges of a graph and to glue them together in a cylindrically
consistent way. In \cite{36a} finally, Hall's construction was generalized
suitably and made applicable to very general phase spaces taking into account
also some dynamical aspects. We will now outline the main idea, following 
\cite{36a} :\\

Central to the subject is the choice of a complex polarization of the
classical phase space. In other words, we must choose the analogue of
$z=x-ip$ of the harmonic oscillator. This is equivalent to choosing a certain
generator $C$ (called complexifier in \cite{36a}) of the associated
complex symplectomorphism which in the case of the harmonic oscillator
consists of the the map $(x,p)\mapsto (z,p)$ and is easily seen to be
$C=p^2/2$ if, as usual, the symplectic structure is defined by
$\{p,x\}=1$. Namely we have
\be \label{3.1}
z=x+i\{x,C\}=\sum_{n=0}^\infty \frac{i^n}{n!} \{x,C\}_n
\ee
where the multiple Poisson bracket is inductively defined by
$\{f,g\}_0=f,\;\{f,g\}_{n+1}=\{\{f,g\}_n,g\}$. It is important for the
existence of the coherent state transform that the polarization is a
positive K\"ahler polarization, in other words, that the generator $C$
is a positive function on the phase space. We will see this in a moment.

The next step consists in the quantization. Following the rule that
Poisson brackets be replaced by commutators times $1/(i\hbar)$ and phase
space functions by operators in a suitable ordering we obtain for the
harmonic ocillator
\be \label{3.2}
\hat{z}=\sum_{n=0}^\infty \frac{\hbar^{-n}}{n!}
[\hat{x},\hat{C}]_n=\hat{W}_t\hat{x}\hat{W}_t^{-1}
\ee
where we have defined
\be \label{3.3}
\hat{W}_t:=e^{-\frac{\hat{C}}{t}}
\ee
where in this case $t=\hbar$ so that $\hat{W}_t=\exp(\hbar\Delta/2)$
where $\Delta=(\partial/\partial x)^2$ is the Laplacian on $\Rl$.
Notice that with our conventions $\hat{p}=i\hbar\partial/\partial x$.
One can check that in the case of the harmonic oscillator this gives
correctly the annihilation operator $\hat{z}=\hat{x}-i\hat{p}$.
We see that the generator $C$ naturally gives rise to the map
$\hat{W}_t$ which due to the positiveness of the operator $\hat{C}$
defines a self-adjoint contraction semi-group of bounded operators.

It is for this reason that the following map, called the kernel of the
coherent state transform, is well-defined
\be \label{3.4}
\rho^t(y,x):=(\hat{W}_t\delta_y)(x)
\ee
which for the harmonic oscillator is easily seen to be the 
standard heat kernel
on $\Rl$, $\delta_y$ being the $\delta$ distribution with respect to
$dx$ supported at $x$.

The coherent states themselves arise as the analytic continuation
of the kernel, that is
\be \label{3.5}
\psi^t_z(x):=\rho^t(y,x)_{y\to z}
\ee
which exists, again, because the operator $\hat{C}$ is positive.
It can be shown for the harmonic oscillator that the naturally arising map
\be \label{3.6}
\hat{U}_t:=\hat{K}\hat{W}_t,
\ee
where $\hat{K}$ denotes analytic continuation, is a unitary map between
${\cal H}=L_2(\Rl,dx)$ and ${\cal H}^\Cl=L_2(\Cl,d\nu_t)\cap\mbox{Hol}(\Cl)$ 
where the latter
denotes the space of square integrable holomorphic functions on $\Cl$
with respect to a measure $\nu_t$ which is constructed from $\rho_t$.
For the case of the harmonic oscillator this latter Hilbert space is
the familiar Bargmann-Segal-Fock space.

In \cite{36b} Hall observed that the case of the harmonic oscillator
can be naturally extended to the case of a cotangent bundle over a connected
compact Lie group $G$, once the following substitutions are made :\\
$\Rl\to G,\; dx\to d\mu_H(h),\;\Cl\to G^\Cl,\; \Delta\to\Delta_G$ where
$G^\Cl$ is the complexification of $G$ and $\Delta_G$ 
denotes the Lalace-Beltrami operator on $G$. In particular, he constructed
the map $\hat{U}_t$ and the measure $\nu_t$. What he did not analyze, 
except for phase space bounds, are the anlytical properties of the states 
$\psi_g(h)$ of (\ref{3.5}), that is, peakedness and Ehrenfest properties. 
Here and in what follows we will always take $h\in G,\;g\in G^\Cl$.

In \cite{36c}, Hall's results were applied to the case of a quantum gauge
field theory. That is, one applies the coherent state transform as generated
by the Laplace Beltrami operator to each copy of the group $G$ associated
with the edges $e$ of a graph of a cylindrical function and obtains a
function cylindrical over the same graph but with the holonomies taking
values in the complexified gauge group. Thus, coherent states become 
functions of $g_e\in G^\Cl,\; e\in E(\gamma)$. While this gives a 
satisfactory mathematical
framework for the construction of measures on $G^\Cl$, the physics
of this map was not understood : namely, not only do we need square
integrable functions on $G^\Cl$ but we also need to know what the complex
connection is which gives rise to the complexified holonomies, that is,
we need to know the map $(A,E)\mapsto A^\Cl$ that expresses the complex
connection as a function of the real phase space. Otherwise, for instance
expectation values which will be functions of the $g_e$ cannot be interpreted
in terms of the $(A,E)$ and thus semi-classical analysis cannot be developed
because, say solutions to the Einstein equations, are formulated 
in terms of the latter.

In order to determine $A^\Cl$ one must determine the classical limit of the
operator which on cylindrical functions reduces to
$\Delta_\gamma:=\sum_{e\in E(\gamma)}\Delta(h_e)$ where $h_e$ is the holonomy
of the real connection of the $G-$bundle along the edge $e$ of $\gamma$.
The problem is, that such a classical limit does not exist !

To see this, notice that \cite{12,13} roughly 
$-\Delta(h_e)\propto(\hat{E}(S_e)_i/\hbar)^2$
where $S_e$ are mutually disjoint analytic surfaces each of which intersects 
the graph only in one point which is an interior point of both $e$ 
and $S_e$ (for definiteness, that intersection can be chosen transversal).
However, it is not possible to write down a {\it single} operator which 
reproduces $\Delta_\gamma$ for {\it every} $\gamma$ {\it and} has a 
classical limit as a well-defined function on the classical phase space
$M$. Namely, suppose first that $\Sigma$ is compact. Since the graph 
$\gamma$ is arbitrary we may consider 
a net of finer and finer graphs $\gamma_\epsilon$ which in the limit 
$\epsilon\to 0$ fill all 
of $\Sigma$. Let us choose the $\gamma_\epsilon$ to be such that 
$\gamma_\epsilon\subset\gamma_{\epsilon'}$ for $\epsilon<\epsilon'$
and to be (subsets of) cubic lattices of spacing $\epsilon$ with 
respect to some spatial background metric. If V is the volume of $\Sigma$ 
as measured by that metric, then in $D$ spatial dimension one will have an 
order of $V/\epsilon^D$ vertices in $\gamma_\epsilon$ each of which accounts 
for 
$D$ surfaces of area of order $\epsilon^{D-1}$. We see that in the classical
limit for sufficiently small $\epsilon$, using the smoothness of the 
classical fields
\be \label{3.0}
\Delta_{\gamma_\epsilon}\to [-\epsilon^{2(D-1)}\sum_{v\in V(\gamma_\epsilon)}
\sum_{I=1}^D [E^a_i(v) n^I_a(v)]^2][1+O(\epsilon)]
\ee
where the sum runs over the vertices of $\gamma_\epsilon$, $n_a^I(v)$
is the normal of the surface $S_{e^I(v)}$ and $e^I(v)$ is an edge
of $\gamma_\epsilon$ that starts at $v$ and runs into the $I$'th  
coordinate direction. This object has a chance to converge in the limit
$\epsilon\to 0$ to a well-defined classical quantity only if 
$2(D-1)=D$, i.e. $D=2$, so that in fact an integral results. For 
$D<2$ this object diverges and for $D>2$ it approaches zero for 
generic field configurations. One could replace $-\Delta_\gamma$ by 
$\sum_{e\in E(\gamma)}
(-\Delta_e)^{D/(2(D-1))}$ in order to fix this (the eigenvalues would still
behave as $j^{D/(D-1)}> j$), however, while this operator now does have
a suitable classical limit at least for the net $\gamma_\epsilon$, 
it is no longer diffeomorphism covariant because it carries the sign of 
the background metric in the definition of the normals $n^I_a(x)$.
If $\Sigma$ is not compact then (\ref{3.0}) 
diverges even in $D=2$ (or its just described replacement in any $D$)
because in gravity the field $E$ does not decay at spatial infinity.

In conclusion, there seems to be no classical limit of the 
cylindrically defined operator $\Delta_\gamma$ as a well-defined, 
diffeomorphism covariant function on $M$ and therefore the interpretation
of the $g_e$ remains obscure. This state of affairs is clearly unsatisfactory 
and there are basically two ways out :\\
Option 1) : One has to choose a different generator of the transform
which {\it actually comes from a well-defined function on $M$}.\\
Option 2) : One gives up the requirement to have a complex continuum
connection $A^\Cl$ altogether and is satisfied with an interpretation of 
$g_e$ in terms of $h_e$ and certain other functions of $A,E$ smeared over
some surfaces $S_e$. Since the latter functions can be interpreted in terms 
of $(A,E)$ one also arrives at an interpretation of $g_e$ and this is
sufficient in order to do semi-classical physics.

In the next two sections we will describe both options in detail.\\
\\
Remark :\\
Before closing this section we would like to point out that a great deal
of properties of the coherent states can be obtained already at this point,
even if the interpretational issue raised above is not yet answered.
Namely, let $\hat{C}_\gamma$ be the cylindrical projections of {\it any}
complexifier and 
\be \label{3.a}
\psi^t_{\gamma,\vec{g}}
:=(e^{-t\hat{C}_\gamma}\delta_{\gamma,\vec{h}})_{|\vec{h}\to\vec{g}}
\ee
where $\vec{g}=\{g^e\}_{e\in E(\gamma)}$ and similar for $\hat{h}$. 
Moreover, define the {\it annihilation and creation operators} 
respectively ($A,B,C,..$ are group indices) 
\be \label{3.b}
\hat{g}^e_{AB}:=e^{-t\hat{C}_\gamma}\hat{h}^e_{AB} e^{t\hat{C}_\gamma}
\mbox{ and }(\hat{g}^e)^\dagger_{AB}
\ee
respectively.
Then, without specifying $\hat{C}_\gamma$ at all, the following 
properties are automatically satisfied (obviously all of this is also
theory independent, in the relations below, with the obvious changes, 
$\vec{h}$ could 
be any configuration coordinates for its cotangent bundle and $\vec{g}$ their
analytical continuations) :
\begin{itemize}
\item[a)] {\it Eigenvalue Property}\\
The coherent states (\ref{3.a}) are eigenstates of any of the 
annihilation operators (\ref{3.b})
\ba \label{3.c}
&&[\hat{g}^e_{AB}\psi^t_{\gamma,\vec{g}}](\vec{h})
=[e^{-t\hat{C}_\gamma}\hat{h}^e_{AB}
\delta_{\gamma,\vec{h}'}](\vec{h})_{|\vec{h}'\to\vec{g}}
\nonumber\\
&=& [e^{-t\hat{C}_\gamma} h^{\prime e}_{AB}
\delta_{\gamma,\vec{h}'}](\vec{h})_{|\vec{h}'\to\vec{g}}
=g^e_{AB}\psi^t_{\vec{g}}(\vec{h})
\ea
simply because the $\delta$-distribution is a generalized eigenfunction
of the multiplication operator in the configuration representation.  
\item[b)] {\it Expectation Values for Normal Ordered Operators}\\
From a) it is trivial to see that 
\be \label{3.d}
\frac{<\psi^t_{\gamma,\vec{g}},P(\{\vec{\hat{g}}^\dagger,\vec{\hat{g}}\})
\psi^t_{\gamma,\vec{g}}>}{||\psi^t_{\gamma,\vec{g}}||^2}
=P(\{\overline{\vec{g}},\vec{g}\})
\ee
where $P$ is any normal ordered polynomial of the creation and annihilation
operators (annihilation operators to the right).
\item[c)] {\it Saturation of the Unquenched Heisenberg Uncertainty 
Relation}\\
Define the symmetric operators
\be \label{3.e}
\hat{x}^e_{AB}:=\frac{1}{2}(\hat{g}^e_{AB}+(\hat{g}^e_{AB})^\dagger),\;
\hat{y}^e_{AB}:=\frac{1}{2i}(\hat{g}^e_{AB}-(\hat{g}^e_{AB})^\dagger)
\ee
then again with a) it is trivial to see that for the fluctuations we find
\ba \label{3.f}
&&\frac{<\psi^t_{\gamma,\vec{g}},(\hat{x}^e_{AB}-x^e_{AB})^2
\psi^t_{\gamma,\vec{g}}>}{||\psi^t_{\gamma,\vec{g}}||^2}
=
\frac{<\psi^t_{\gamma,\vec{g}},(\hat{y}^e_{AB}-y^e_{AB})^2
\psi^t_{\gamma,\vec{g}}>}{||\psi^t_{\gamma,\vec{g}}||^2}
\nonumber\\
&=&\frac{1}{2}
\frac{|<\psi^t_{\gamma,\vec{g}},[\hat{x}^e_{AB},\hat{y}^e_{AB}]
\psi^t_{\gamma,\vec{g}}>|}{||\psi^t_{\gamma,\vec{g}}||^2}
\ea
\item[c)'] {\it Reproducing Property}\\
The connection between the coherent state transform 
$\hat{U}^t_\gamma:\;{\cal H}_\gamma\mapsto {\cal H}^\Cl_\gamma$,
defined analogously to (\ref{3.6}), and the coherent states is summarized by 
the following reproducing property, valid for any $\psi\in {\cal H}_\gamma$ :
\be \label{3.g}
(\hat{U}_t\psi)(\vec{g})=<\psi^t_{\gamma,\vec{g}^\ast},\psi>
\ee
where $g\mapsto g^\ast$ is the unique involution on $G^\Cl$ that preserves
$G$ (this formula can be proved by using the expansion of the group 
$\delta$ distribution in terms of characters according to the Peter\&Weyl
theorem, see e.g. \cite{31}). 
\end{itemize}
The additional properties that one would like the coherent states to possess
and which do not directly follow from the general form (\ref{3.a}) are 
the following, for which we need now the expression for $g_e$ in terms of 
$A,E$ :
\begin{itemize}
\item[d)] {\it Peakedness Properties}\\
We want the coherent states (\ref{3.a}) to be peaked in the configuration 
representation at $A$, in the momentum representation at $E$ and in the 
Bargmann-Segal representation ${\cal H}^\Cl$ (the image of $\cal H$ under 
$\hat{U}_t$
to be defined for general $\hat{C}$ along the lines outlined in \cite{36a})
at $(A,E)$. For instance, if with respect to $\gamma$ we take as 
elementary configuration coordinates the holonomies $h_e$ and as elementary
momentum coordinates the $E_i(S_e)$ considered above and if we know the 
explicit formula $g_e(\{h_{e'},E(S_{e'})\})$ which is supposed to 
be invertible, then we want the probability amplitudes for the coherent state
with label $\vec{g}$ in the configuration --, momentum -- and 
Segal-Bargmann Hilbert spaces respectively to be peaked at 
$h_e(\vec{g}),\;[E(S_e)](\vec{g}),\;\vec{g}$ respectively.
Notice that if we take $\vec{g}\in G^{|E(\gamma)|}$ then as $t\to 0$
$\psi^t_{\gamma,\vec{g}}(\vec{h})$ on $G^{|E(\gamma)|}$ is supported at
$\vec{g}=\vec{h}$ for any choice of complexifier $\hat{C}_\gamma$ since
by its very definition $\psi^t_{\gamma,\vec{g}}$ approaches 
$\delta(\vec{g},\vec{h})$ as $t\to 0$.
\item[e)] {\it Ehrenfest Property}\\
While expectation values of normal ordered polynomials of alternation
operators already have the correct expectation values without 
quantum corrections, we want that to leading order in $t$ or $\hbar$ also 
the elementary operators associated with $h_e, E(S_e)$ as well as their 
various commutators divided by $i\hbar$ have the correct expectation value
guaranteeing the correct infinitesimal quantum dynamics. The fact that
the alternation operators do have the correct expectation values makes
it plausible that also this property can be verified for any 
$\hat{C}_\gamma$.
\item[f)] {\it Overcompleteness}\\
The coherent states should be overcomplete in order to be able to 
approximate any possible physical situation. Overcompleteness follows 
automatically if the coherent state transform $\hat{U}_t:\;{\cal H}\mapsto 
{\cal H}^\Cl$ is unitary since then that map is onto. More precisely, 
due to the reproducing property (see e.g. \cite{31}) :
\be \label{3.g1}
1_{{\cal H}_\gamma}=\int_{(G^\Cl)^{|E(\gamma)|}}d\nu_t(\vec{g})
|\psi^t_{\gamma,\vec{g}^\ast}><\psi^t_{\gamma,\vec{g}^\ast}|
\ee
A method for a constructive proof for general $\hat{C}$, up to analytical
details, is given in \cite{36a}. Namely, the measure $\nu_t$ can be uniquely
determined if the operator $\hat{W}_t$ is well-defined and if the
cylindrical family of measures constructed in \cite{36a} can be extended
to a $\sigma-$additive measure on the projective limit of the cylindrical
projections of spaces of complex quantum connections that one can define
in analogy to \cite{36c}.\\
Overcompleteness is actually also rather plausible for general $\hat{C}$
by inspection
because these states arise as the ``evolution'' under $\hat{W}_t$ of the 
$\delta$ distributions. Now the latter provide a complete basis of 
generalized functions and $\hat{W}_t$ is invertible on a dense domain 
of $\hat{W}_t^{-1}$ (the inverse is certainly not bounded). 
\item[h)] {\it Diffeomorphism Covariance}\\
The coherent states should, as all the other states of the Hilbert space,
transform covariantly under the diffeomorphism group.
This will be the case 
provided that the operator $\hat{C}_\gamma$ is itself diffeomorphism 
covariant (does not make use of any background structure), specifically,
$\hat{U}(\varphi)\hat{C}_\gamma\hat{U}(\varphi)^{-1}=
\hat{C}_{\varphi^{-1}(\gamma)}$
where $\mbox{Diff}(\Sigma)\ni\varphi\mapsto\hat{U}(\varphi)$ is the unitary
representation of the diffeomorphism group described in \cite{7}.
\end{itemize}

\subsection{Option 1) : The Volume Operator as the Complexifier}
\label{s3.2}

In this section we modify the coherent state transform by choosing a 
different complexifier. We will argue now that (a suitable power of)
the ``volume'' of a region $R\subset \Sigma$
\be \label{3.7}
V(R):=\int_R d^Dx \sqrt{\det(q)(x)}
\ee
is the most natural candidate. In case that $\Sigma$ is compact or that
classically the fields vanish sufficiently fast at spatial infinity
as in Yang-Mills theory, we will take $R=\Sigma$ in the sequel. Otherwise,
we will take $R$ to be a bounded region to begin with and send $R\to \Sigma$
only after all calculations have been performed.
Here, 
\be \label{3.8}
\det(q):=\sqrt[D-1]{\det(E^a_i E^b_i)}
\ee
and (\ref{3.7}) is called the
volume functional because in the case of general relativity
$E^a_i=\sqrt{\det(q)}e^a_i$ where $e^a_i$ is the $D$-bein field and
$q_{ab}$ is the $D$-metric intrinsic to $\Sigma$.

The reasons are as follows :\\
(i)\\
As it is clear from the discussion in the previous section, it is
important that $C$ is a positive semi-definite function on the phase space
as this translates into a positive definite operator upon quantization.
The volume has this property.\\
(ii)\\
Notice that even in the case of gauge theories on a background metric
the electric field is a Lie algebra valued vector density of weight one.
Therefore, $E^a_i E^b_i=\det(q) q^{ab}$ is in general a gauge invariant
tensor density of weight two. Hence, (\ref{3.8}) is a scalar density of
weight two which can be constructed without any background structure and 
therefore the volume functional is {\it diffeomorphism invariant} if
$R=\Sigma$ and {\it diffeomorphism covariant} if $R\subset\Sigma$ !
This is important in order to obtain diffeomorphism covariant coherent
states in the case of diffeomorphim invariant quantum field theories of
connections.\\
(iii)\\
As we want to start with a Hilbert space which consists of square integrable
functions of connections for which the connection operator is a
multiplication operator, it is natural to consider an operator which is
entirely constructed from the electric field operator so that the
analogue of $z$ is given by $Z_a^j=A_a^j+i f_a^j(E)$. The volume density is
the simplest scalar density of weight one entirely constructed from
electric fields.\\
(iv)\\
Using the symplectic structure $\{A_a^i(x),E^b_j(y)\}=-\kappa\delta_a^b
\delta^i_j \delta(x,y)$ where $\kappa$ is the coupling constant
and
\be \label{3.9}
C(R):=\frac{1}{\lambda\kappa} V(R)^n
\ee
where $n\ge 1$ is a positive real number and
$\lambda$ is a positive, possibly dimensionful, parameter
so chosen that for $x\in R$
\be \label{3.10}
f_a^j(x):=\{C(R),A_a^j(x)\}=\frac{n V(R)^{n-1}}{\lambda}
\frac{\partial \root 2(D-1)
\of{\det(E^b_i E^c_i)}}{\partial E^a_j}=:n V(R)^{n-1}
\frac{e_a^i}{\lambda(D-1)}
\ee
has dimension of inverse length we easily see that $E^a_j$ can be
reconstructed from $f_a^j$ and therefore the complex connection
$Z_a^j$ together with its complex conjuate contains full phase space 
information. The field $e_a^i$ is the co-$D$-bein in general relativity.

Notice that $Z_a^j=A_a^j-if_a^j$ really transforms as a $G-$connection under
gauge transformations since $\delta Z=-d\Lambda+[\Lambda,A]+i[\Lambda,e]
=-d\Lambda+i[\Lambda,Z]$ so that the coherent state transform is both
diffeomorphism covariant and gauge covariant.\\
(v)\\
Finally, to be useful, it is necessary that one can quantize the generator.
But this is the case for the volume functional in any dimension along the
lines of \cite{12,14,15,16}. Moreover, on the Hilbert space that we have 
chosen
in section \ref{s2} the spectrum of that operator is entirely discrete and,
although very complicated, explicitly known at least in terms of matrix 
elements \cite{16,8}. \\
\\
Upon quantization $\hat{E}^a_i=i\hbar\kappa\delta/\delta A_a^i$
and the generator takes the following form on cylindrical functions 
\be \label{3.11}
\hat{C}(R)=\frac{(\hbar\kappa)^{n\frac{D}{D-1}}}{\lambda\kappa} \hat{v}
\ee
where $\hat{v}$ is a dimensionless operator constructed from invariant
vector fields corresponding to the copies of the group associated with the
edges of graphs. The coherent state transform is then generated by
\be \label{3.12}
\hat{W}_t=e^{-t\hat{v}} \mbox{ where }
t=\frac{(\hbar\kappa)^{n\frac{D}{D-1}-1}}{\lambda}
\ee
is a dimensionless parameter which vanishes as $\hbar\to 0$. For instance,
for general relativity in $3+1$ dimensions, $\hbar\kappa$ is the Planck area.
\\
\\
Next, we define coherent states in analogy to (\ref{3.5}). The idea
is to define coherent states graphwise, which means that the state
approximates a certain point in the classical phase space {\it on that
graph only}. We can do this for every graph which is contained in the
region $R$. In case that $R\not=\Sigma$ this does not exclude the
possibility to have graphs which run to spatial infinity :
We can use the asymptotic structure available and allow only such graphs
which run to spatial infinity inside fixed ``thin tubes" of $R$ which have
finite Lebesgue measure. Notice that these complications would not be
necessary if we would choose $n=1$. In general we cannot choose $n=1$ for 
reasons explained below, see also the model described in section \ref{s6}.\\

The fundamental definition is
\be \label{3.13}
\psi^t_{\gamma,Z}(A):=(\hat{W}_t\delta_{\gamma,A'})(A)_{|A'\to Z}
\ee
where the $\delta$ distribution in (\ref{3.13}) is defined by
\be \label{3.14}
\delta_{\gamma,A'}(A):=\sum_{\vec{j},\vec{J}}
\overline{T_{\gamma,\vec{j},\vec{J}}(A)}
T_{\gamma,\vec{j},\vec{J}}(A')
\ee
and where the sum is over all possible not necessarily gauge invariant
spin-network functions on that graph $\gamma$ if we work at the
non-gauge invariant level while it is over all possible
gauge invariant functions only if we work at the gauge invariant level.
It is important to stress that in (\ref{3.14}) we include only
spin-network states whose vector of representations $\vec{j}$ does
not contain a zero entry.

We thus obtain coherent states $\psi^t_{\gamma,Z}$ which have the property
to be orthogonal,\\ $<\psi^t_{\gamma,Z},\psi^t_{\gamma',Z'}>=0$, if
their underlying graphs are different, $\gamma\not=\gamma'$.
We also define coherent states of a different type,
\be \label{3.14a}
\Psi^t_{\gamma,Z}:=\sum_{\gamma'\subset\gamma}\psi^t_{\gamma',Z}
\ee
where the sum extends over all subgraphs of 
$\gamma$ which can
be obtained from $\gamma$ by deleting edges of $\gamma$ in all possible
ways (if the state is to be gauge invariant then the sum extends over
closed subgraphs only). The idea is 
not to take inner products of states (\ref{3.14a}) with different $\gamma$
but only between those with the same $\gamma$ but different $Z,Z'$.
In other words, one first restricts the Hilbert space $\cal H$
to the completion ${\cal H}_\gamma$ of the span of spin-network states
over closed subgraphs of $\gamma$ and then one lets $\gamma$ grow.
Recall that given two piecewise analytic graphs, their union is still
a piecewise analytic graph. (We cannot immediately transfer our definitions
to the smooth category of webs \cite{40} because there we do not have
the notion of an orthogonal basis unless we restrict ourselves to
non-degenerate webs \cite{41}; we will, however, not go into this subject in
the present paper). Therefore, there is a generalized projective structure on
the set of piecewise analytical graphs and 
the final coherent state $\Psi^t_Z$ is a {\it generalized 
projective limit} of the states $\Psi^t_{\gamma,Z}$. 

Let us illustrate the situation by drawing an analogy with the coherent
states for,
say, an (in)finite number $N\le\infty$ of harmonic oscillators :
The role of the graph label
$\gamma=(e_1,..,e_n),\;n<\infty$, given as a finite collection of edges,
is played by the mode label $\vec{k}=(k_1,..,k_n),\;n<N$, given by a
finite collection of non-negative integers. The analogues of the states
$\Psi^t_{\gamma,Z}=\Psi^t_{\gamma,\vec{g}}$ with
$\vec{g}=(h_{e_1}(Z),..,h_{e_n}(Z))$ is given by the coherent state
for $n$ uncoupled harmonic oscillators $\Psi^t_{\vec{k},\vec{z}}$
with an array of complex numbers $\vec{z}=(z_{k_1},..,z_{k_n})$
corresponding to the mode vector $\vec{k}$. The projective limit
of taking the ``biggest possible graph'' corresponds to taking the
(in)finite direct product limit
$\vec{k}\to(1,2,..,N)$ and one obtains the full coherent state
$\Psi^t_Z,\;Z=(z_1,z_2,..,z_N)$. One does not compute inner products
between
states with different $\vec{k}$ but only with different $\vec{z}$ for the
same $\vec{k}$ which models the properties of $\Psi^t_Z$ on its
cylindrical projections $\Psi^t_{\vec{k},\vec{z}}$. The analogues
of the states $\psi^t_{\gamma,Z}$ are the states
$\psi^t_{\vec{k},\vec{z}}=\Psi^t_{\vec{k},\vec{z}}-
\Omega<\Omega,\Psi^t_{\vec{k},\vec{z}}$ 
where we have taken out the vacuum mode so that
$<\psi^t_{\vec{k},\vec{z}},\psi^t_{\vec{k}',\vec{z}'}>=0$ for
$\vec{k}\not=\vec{k}'$.

Notice that the restriction of $\Psi^t_{\gamma,Z}(A)$ to real valued
$Z=A'$ is the ``heat kernel" $\rho_{\gamma,t}(A,A')$ for the ``heat
equation"
\be \label{3.15}
[\partial/\partial t+\hat{v}]\rho_{\gamma,t}(A,A')=0\mbox{ such that }
\rho_{\gamma,0}(A,A')=\delta_\gamma(A,A')\;.
\ee
As the volume operator is an essentially self-adjoint, positive semi-definite
operator with discrete spectrum which leaves the subspace of $\cal H$
spanned by spin-network states of given $\gamma,\vec{j}$ invariant, we can
diagonalize it and define
another orthonormal basis of eigenstates $T_{\gamma,\lambda,n}$
of $\hat{v}$ where $\lambda$ labels the eigenvalue and the integer $n$
its degeneracy. We can then write (\ref{3.14}) alternatively as
\be \label{3.16}
\delta_{\gamma,A}(A'):=\sum_{\lambda,n}
\overline{T_{\gamma,\lambda,n}(A')}
T_{\gamma,\lambda,n}(A)
\ee
which allows us to explictly compute the coherent states as
\be \label{3.17}
\psi_{Z,\gamma,t}(A)=\sum_{\lambda,n}e^{-t\lambda}
T_{\gamma,\lambda,n}(Z)
\overline{T_{\gamma,\lambda,n}(A)}\;.
\ee
The function (\ref{3.17}) is to be understood in the following sense :
Given a point in the classical phase space $A,E$, compute the $G^\Cl$
connection $Z=A-if(E)$ and from this its holonomies $h_e^\Cl:=h_e(Z)$
for each edge $e$ of $\gamma$. Then insert these elments of $G^\Cl$ into
the eigenfunctions appearing in the series in (\ref{3.17}).

Several points of worry arise when looking at (\ref{3.17}) :\\
(i)\\
Does the series in (\ref{3.17}) converge, in the sup-norm topology
with respect to $G^n$,
where $n$ denotes the number of edges of $\gamma$ ? This will,
in particular, not
be the case if one of the $\lambda$ has infinite multiplicity. The volume
operator as defined in \cite{12,14,15,16}, however, has presumably
precisely this property for the zero eigenvalue, at least in the case of
general relativity in 3+1 dimensions which requires $G=SU(2)$! Thus, in
this case, in order to make
sense of (\ref{3.17}) we must discard the zero volume eigenstates even
from the kinematical Hilbert space. (In particular this has to be done
at the gauge non-invariant level). This is quite satisfactory because
the classical phase space can be viewed as a cotangent bundle over
smooth, signature $(+,..,+)$ $D$-metrics for which vanishing volume,
that is, vanishing determinant of the three-metric, is not allowed.
That the signature is $(+,..,+)$ is guaranteed if we restrict to states
with non-vanishing expectation value for the area operator
\cite{12,13} for every surface that intersects the graph.

But even if all eigenvalues have finite multiplicity, the series does not
necessarily converge : while $T_{\gamma,\lambda,n}$ is a bounded function
of $G^n$, it is not any longer so of $(G^\Cl)^n$ because that group is
not compact. What is needed, roughly speaking, is the following :\\
we can decompose the $T_{\gamma,\lambda,n}$ in terms of spin-network
functions which turns the above series into a series over $\vec{j},
\vec{J}$. The coefficient of $T_{\gamma,\vec{j},\vec{J}}(Z)$ is of the
form $e^{-t\lambda(\vec{j},\vec{J})}$ times something that grows at most
linearly with $\vec{j},\vec{J}$ while $T_{\gamma,\vec{j},\vec{J}}(Z)$
grows exponentially with $\vec{j},\vec{J}$ for $Z$ in the non-compact
directions of $G^\Cl$. Thus, for the series to converge it would be
sufficient if
\be \label{3.18} 
\lambda(\vec{j},\vec{J})\ge c(\sum_{e\in E(\gamma)} j_e+\sum_{v\in V(\gamma}
J_v)^{1+\epsilon}
\ee
where $c$ is a positive number independent of $\vec{j},\vec{J}$ and
$\epsilon$ can be any positive number. The criterion (\ref{3.18}) is a
condition on the spectrum of $\hat{v}$ which needs to be checked to hold.
This is the reason why we have allowed for a power $n$ different from
$n=1$ in (\ref{3.9}) : by taking $n$ sufficiently large we can guarantee
that criteron (\ref{3.18}) is satisfied.

More precisely we have the following :\\
Looking at (\ref{3.7}), (\ref{3.9}) and the explicit expression for the
volume operator as derived in \cite{12,14,15,16} we infer that the electric
fields get, roughly speaking, replaced by right invariant vector fields
$X^i_e:=X^i(h_e)$ on the various copies of $G$ corresponding to the edges of
$\gamma$. As those act on spin-network functions roughly by multiplication
by $j_e$, we find the eigenvalues of the volume opertor to be of the
form
\be \label{3.19}
\lambda(\vec{j},\vec{J})=(P_{2D}(\vec{j},\vec{J})^{n/(2(D-1))}
\ee
where $P_{2D}$ is a homogenous, positive polynomial of degree $2D$
which depends non-trivially on all the variables $\vec{j},\vec{J}$.
Obviously, taking $n>2(D-1)$ we have good chances to satisfy (\ref{3.18}).
Presumably,
$n>D-1$ will be sufficient since because of gauge invariance the
$\vec{j},\vec{J}$ do not have independent ranges.
For instance, for $SU(2)$, due to gauge
invariance the sum of all but one, say $j_{e_0}$, of those $j_e$ that
correspond to $e$'s which meet at a common vertex must always exceed the
value of $j_{e_0}$. Moreover, the value of $J_v$ is bounded by the sum of
all those $j_e$. These relations hold for all of the vertices
and thus there is a good chance that
we can estimate (\ref{3.19}) as
\be \label{3.20}
\lambda(\vec{j},\vec{J})\ge c(\max(\vec{j},\vec{J}))^{\frac{n}{D-1}}
\ee
which would be sufficient for any $D$. However, this must be checked
in the case at hand. In particular, it could happen that the function
$\lambda(\vec{j},\vec{J})$ has ``degenerate directions" in which case
even large $n$ would not help to make the series converge. \\
(ii)\\
Even if the series converges, are these coherent states square integrable ?
We easily see that the convergence of the series is sufficient for this to be
the case. Namely, if the series converges, we compute the norm as
\be \label{3.21}
||\psi_Z||^2=\sum_{\lambda,n} e^{-2t\lambda}|T_{\gamma,\lambda,n}(Z)|^2
\ee
which then certainly converges as well.\\
(iii)\\
Finally, is the generalized projective limit $\Psi^t_Z$ of the states
$\Psi^t_{\gamma,Z}$ square integrable ? There is no, not even partial
aswer to this question available at the moment, however, notice that
even the uncountably infinite direct product limit of an uncountably
infinite number of harmonic oscillators is square integrable. This
follows immediately from the Kolmogorov theorem \cite{45} for
an uncountably infinite tensor product of probability measure (here :
Gaussian measures) Hilbert spaces. Thus, the normalizability of
$\Psi^t_Z$ is indeed conceivable.\\
(iv)\\
If we can then verify the properties (a)-(h) mentioned above, what we will
have achieved is that we have states that are peaked on a classical
configuration $Z$ in the sense that the operator $\hat{g}^e$ 
corresponding
to $g^e=h^e(Z)$ has expectation value $g^e$, saturates the
Heisenberg uncertainty bound etc. However, since all these properties
(a)-(h) are verified for $g^e$ only, we must ask whether
we can reconstruct $Z$ from all the $h^e(Z)$, that is, whether
the holonomies separate the points on the space of smooth complexified
connections.

This is a non-trivial question due to the presence of so-called
null-rotations for non-compact gauge groups and 
amounts to proving a Giles' theorem \cite{46} for non-compact
gauge groups. At least for $SU(2)^\Cl=SL(2,\Cl)$, this has been answered
affirmatively in an appropriate sense in \cite{47} and we believe the proof 
to be valid generally
for complexifications of compact connected gauge groups. If we work
at the gauge non-invariant level, the proof is obvious since we just
have to consider the limit of infinitesimal open paths.\\
\\
\\
We now argue that the coherent states (\ref{3.14a})
so constructed have very good chances to satisfy all the
properties (d)-(h) mentioned above, assuming that there are no convergence
problems even at the gauge non-invariant level. We will indicate the
necessary modifications of the analysis when we restrict to the
gauge invariant sector. The analysis is in fact quite general and can
be generalized to the quantization of any field theory with a 
generalized projective
structure, once a choice of the complexifier $C$ and a choice of polarization
of the classical phase space has been made. \\
(d)\\
The way in which these states are localized is obscure at the moment.
In this paper we will just outline how one might prove this property.
First of all, notice that the coherent states become, for real connections
$Z$, just $\delta$ distributions on $\gamma$ in the semi-classical limit
as $\hbar\to 0$ (that is, $t\to 0$, see (\ref{3.12}), (\ref{3.15})).
Thus, in the connection representation, the state $|Z,\gamma,t>$
is certainly peaked at $A=\Re(Z)$ as $t\to 0$ for $\Im(Z)=0$ for any
$\gamma$. What happens if $\Im(Z)\not=0$ is unclear at the moment.
Next, we want to study the state $|Z,\gamma,t>$ in the momentum or
electric field representation which is nothing else than the spin-network
representation (see \cite{16}).
Now, the representation (\ref{3.17}), with the $T_{\gamma,\lambda,n}$
written in the spin-network basis, is not immediately useful in order
to study the behaviour of the state in the limit $t\to 0$ because
the exponential terms become unity in the limit $t\to 0$, that is, the 
convergence of the series worsens in the limit $t\to 0$. The idea is to
use a Poisson summation formula which exists for all compact gauge groups
\cite{48,36b} and which should transform
the series into a series with coefficients of the form
$\exp(-\lambda(\vec{j},\vec{J})/t^\alpha)$ where $\alpha$ is a positive
number. In the limit $t\to 0$ then the leading term would be the one with
$\lambda$ closest to zero and this would be the peakedness property in the
electric field representation. We will actually use this method 
in the next paper of this series \cite{31} for the original heat kernel
complexifier. \\
(e)\\
To prove the Ehrenfest property is very much like proving the peakedness 
property in the Bargmann Segal representation and also should be based on
the Poisson summation formula. The reason is that expectation values of 
polynomials in the basic operators can be expanded, using overcompleteness
of the coherent states, as a polynomial in the matrix elements between 
normalized coherent states $\xi^t_{\gamma,\vec{g}}$ where the extra 
variables $\vec{g}'$ as in (\ref{3.g1}) are integrated over with respect to
$d\nu_t(\vec{g}')||\psi^t_{\gamma,\vec{g}}||^2$. But then the Ehrenfest 
property follows once we find for any elementary operator $\hat{O}$ that 
$$
<\xi^t_{\gamma,\vec{g}},\hat{O}\xi^t_{\gamma,\vec{g}'}>
=O(\vec{g})
<\xi^t_{\gamma,\vec{g}},\xi^t_{\gamma,\vec{g}'}>(1+O(t))
$$
which in turn should be easy to establish
if the overlap function on the right hand side of this equality is 
peaked at $\vec{g}=\vec{g}'$. But the latter property is just the same 
as the peakedness property in the Segal-Bargmann representation which
can be seen generally from the reproducing property.\\ 
(f)\\
As already said, the (over)completeness of the coherent states in the
kinematical Hilbert space ${\cal H}=L_2(\overline{{\cal A}},d\mu_0)$
would follow trivially if one could
establish that the map (\ref{3.6}), generalized to our context, is a
unitary map between $\cal H$ and a suitable $L_2$ space of holomorphic
functions of complex connections with respect to a measure $\nu_t$ because
then the map $\hat{W}_t$ would be onto, in particular. In addition, the 
general comments from the previous section apply.\\
(h)\\
The coherent states of this section are diffeomorphism covariant by their 
very construction.\\ \\
This concludes the general outline of how one might construct coherent
states for quantum gauge theories from a coherent state transform which
can also be interpreted in terms of a complex connection $A^\Cl$. In 
\cite{31} we will, however, not use the volume operator as the complexifier
for the following reasons :\\
i)\\
The spectrum of the volume operator is not explicitly known. This lack
of knowledge makes analytical proofs very hard although a numerical 
method is of course possible.\\
ii)\\
More serious is the following observation : Unless $V(R)$ itself
is a polynomial function of the $E_i(S)$, then even classically the 
$g^e_{AB},\bar{g}^e_{AB}$ do not a form a $^\ast$ Poisson algebra for 
$D>2$. This becomes obvious from the fact that while
$\{A_a^{j\Cl}(x),A_b^{k\Cl}(y)\}=\{E^a_j(x),E^b_k(y)\}=0,
\{A_a^{j\Cl}(x),E^b_k(y)\}=-\delta_a^b\delta^j_k\delta(x,y)$
(the complexifier induces a canonical transformation) we have
$$
\{A_a^{j\Cl}(x),\overline{A_b^{k\Cl}(y)}\}=
\delta(x,y)\frac{\partial^2 V(R)}{\partial E^a_j(x) \partial E^b_k(x)} 
+\mbox{more}
$$
where ``more'' is non-distributional. 
Thus, since the connections are only smeared in one spatial direction 
inside a holonomy functional, it follows that for $D>2$ the Poisson bracket
$\{g_{AB}(A^\Cl),\overline{g_{AB}(A^\Cl)}\}$ is necessarily distributional
or even ill-defined and does not lie in the original Poisson algebra 
any longer. This means that the fluctutions of the $\hat{x}_{AB}, 
\hat{y}_{AB}$ are ill-defined if the Ehrenfest property holds because 
the right hand side of (\ref{3.f}) will then be proportional to
$\{g_{AB}(A^\Cl),\overline{g_{AB}(A^\Cl)}\}$ to first order in $t$.
Whether or not this is bad is unclear, after all it is unnecessary to 
work with $\hat{g}_{AB}$ itself. On the other hand, due to the eigenvalue 
property 
and the similarity with the creation and annihilation operator algebra 
it would be very convenient to have the $\hat{g}_{AB}$ at one's disposal.\\
\\
Due to these difficulties we will turn to option ii) in the remainder 
of this paper and the subsequent issues of this series. It should be kept
in mind, however, that option 1) exists. Its obvious advantage is that 
one has an actual complex connection which implies that one can work
entirely with graphs and never needs the additional dual polyhedronal
decompositions which are a source of ambiguity.

\subsection{Option 2) : The Heat Kernel Complexifier}
\label{s3.3}

In this section we will be satisfied with obtaining $g_e$ as a definite
function of the functions $h_e,P_e$ described in section \ref{s2.1}.
We do not require that $g_e$ is itself the holonomy along $e$ for 
some complex connection $A^\Cl$.

The results of this section hold for arbitrary compact, semisimple connected 
gauge groups and direct products of such with Abelian ones.

As we want to bring in Planck's constant $\hbar$ as a measure of closeness
to classical physics, we need to spend a few moments on dimensionalities
as in the previous section for the volume functional.
The dimension of the time coordinate $x^0$ 
is taken to be the same as that of the spatial coordinates $x^a$,
namely $[x^0]=[x^a]=$cm$^1$ which can always be achieved by absorbing 
an appropriate power of the speed of light into the coupling constant 
$\kappa$ of the theory.

We will take our connection one-form to be of dimension $[A]=$cm$^{-1}$
so that its holonomy is dimensionless. In $D+1$ spacetime dimensions
the kinetic term of the classical action is given by
$$
A_{kin}=\frac{1}{\kappa}\int_\Rl dt\int_{\Sigma} d^Dx\; E^a_i(x) 
\dot{A}_a^i(x) $$
and its dimension is that of an action, that is, $[A_{kin}]=[\hbar]$. 
In Yang-Mills theories the electric field is a first derivative of 
$A_a^i$ and thus has dimension $[E^a_i]=$cm$^{-2}$. In general relativity
the metric components, the D-beins and also $[E^a_i]=$cm$^0$ are 
dimensionfree. It follows that in Yang-Mills (YM) theory the Feinstruktur
constant 
\be \label{8.1}
\alpha:=\hbar\kappa 
\ee
has dimension $[\alpha]:=$cm$^{D-3}$ and in general relativity (GR) 
$[\alpha]=$cm$^{D-1}$. 

Let now $\gamma$ be a graph and consider the symplectic manifold
$(M_\gamma,\Omega_\gamma)$ introduced in section \ref{2.1}
with its canonical coordinates $h_e,P^e_i:\;e\in E(\gamma)$. 
The electric flux variable (\ref{2.5}) then 
has dimension $[P^e_i]=$cm$^{D-3}$ in YM and cm$^{D-1}$ in GR respectively
and in general let $[P^e_i]=$cm$^{n'_D}$.
Let now $a$ be an arbitrary but fixed constant with the dimension of a 
length, $[a]=$cm$^1$, say $a=1$cm if $n_D\not=0$ and let $a$ be 
dimensionfree otherwise. Then we introduce the dimensionfree
quantity
\be \label{8.2}
p^e_i:=\frac{P^e_i}{a^{n_D}} 
\ee
where $n_D=n'_D$ if $n'_D\not=0$ and $n_D=1$ otherwise.
Notice that a natural choice for a dimensionful constant 
in general relativity in any $D>1$ would
be $a=1/\sqrt{|\Lambda|}$ where $\Lambda$ is the (supposed to be 
non-vanishing) cosmological constant.

On the other hand, it is $E^a_i/\kappa$ which is canonically conjugate 
to $A_a^i$ rather than $E^a_i$ itself, therefore the brackets
(\ref{2.6}) get modified into
\ba \label{8.3}
\{h_e,h_{e'}\}_\gamma &=& 0\nonumber\\
\{\frac{P^e_i}{\kappa},h_{e'}\}_\gamma &=&
\delta^e_{e'} \frac{\tau_i}{2}h_e\nonumber\\
\{\frac{P^e_i}{\kappa},\frac{P^{e'}_j}{\kappa}\}_\gamma &=&
-\delta^{ee'}f_{ij}\;^k \frac{P^e_k}{\kappa}
\ea
We are now ready to define the complexifier for the symplectic manifold 
$M_\gamma$, it is given by
\be \label{8.4}
C_\gamma:=\frac{1}{2\kappa a^{n_D}}\sum_{e\in E(\gamma)}\delta^{ij}
P^e_i P^e_j
\ee
and since $C_\gamma$ is gauge invariant it will pass to the reduced phase 
space. Using the partial order $\prec$ of \cite{37} or section 
\ref{s2.1} it is immediately
clear that $C_\gamma$ defines a self-consistently defined function on
the $M_\gamma$, that is, for $\gamma\prec\gamma'$ we have 
$\{p_{\gamma'\gamma}^\ast C_\gamma, p_{\gamma'\gamma}^\ast 
f_\gamma\}_{\gamma'}=p_{\gamma'\gamma}^\ast \{C_\gamma,f_\gamma\}_\gamma $
for any $f_\gamma\in C^\infty(M_\gamma)$.

We can explicitly compute the complexified holonomy and complexified 
momenta for any compact, semi-simple gauge group $G$. Since 
$\{P^e_i,C_\gamma\}=0$ (gauge invariance of $C_\gamma$) we have 
\ba \label{8.5}
\{h_e,C_\gamma\}_\gamma &=& -P^e_i\frac{\tau_i}{2 a^{n_D}} 
h_e=-p^e_i\frac{\tau_i}{2} h_e 
\nonumber\\ 
\{h_e,C_\gamma\}_{\gamma(2)}&=& \frac{1}{a^{2n_D}}P^e_i P^e_j
\frac{\tau_i\tau_j}{4} h_e
=(-p^e_j\frac{\tau_j}{2})^2 h_e 
\ea 
where we define generally $p^e:=\sqrt{p^e_j p^e_j}$.
In the second line of (\ref{8.5}) we have made use of the fact that $G$ is 
semi-simple so that the structure constants are completely skew and so 
$\{p^e_j,C_\gamma\}=0$.

We therefore conclude that the complexification of 
$h_e$ is given by 
\ba \label{8.6}
h^\Cl_e &:=& g_e =\sum_{n=0}^\infty \frac{i^n}{n!}\{h_e,C\}_{(n)}
\nonumber\\
&=& [\sum_{n=0}^\infty \frac{i^n}{n!} (-p^e_j\frac{\tau_j}{2})^n] h_e 
\nonumber\\
&=& e^{-i\tau_j p^e_j/2} h_e=:H_e h_e 
\ea
and similarly $P^{e\Cl}_i=P^e_i$. Thus we have established the following.
\begin{Lemma} \label{la3.1}~\\
The complexification of the holonomy for compact and semisimple $G$ is given 
directly as a left 
polar decomposition, where the right unitary factor is the holonomy of the 
compact gauge group while the left positive definite hermitean factor is
just the exponential of $-i p^e_j\tau_j/2$.
\end{Lemma}
For $G=U(1)$ the generator $\tau_j/2$ has to be 
replaced by the imaginary unit $i$.

Notice that (\ref{8.6}) makes sense since $p^e_j$ is dimensionless. 
Moreover, we have naturally stumbled on the diffeomorphism 
\cite{36b} 
\be \label{8.7}
\Phi\; :\; T^\ast(G)\mapsto G^\Cl;\; (p^j,h)\to g:=Hh=e^{-ip^j\tau_j/2}h\;.
\ee
The diffeomorphism (\ref{8.7}) has a further consequence : 
$(T^\ast(G),\omega)$ is a symplectic manifold while $G^\Cl$ is a complex 
manifold. Thus, $T^\ast(G)$ is a symplectic manifold with a complex structure
which, as one can show (\cite{36b,37} and references therein), is 
$\omega$-compatible. In fact, $\omega$ is just given by (\ref{8.3}) 
with $P^e_i$ replaced by $p_i$ and the label $e=e'$ dropped.
Therefore, $T^\ast(G)$ is in fact a K\"ahler 
manifold and a Segal-Bargmann representation (wave functions depending on 
$g$) corresponds to a positive K\"ahler polarization \cite{53}.

Finally, let us compute the Segal-Bargmann transform corresponding to 
$C_\gamma$ as in \cite{36a}. As follows from the previous 
section, we have in 
the connection representation (wave functions depending on the $h_e$)
\be \label{8.8}
\hat{P}^e_j=\frac{i\hbar\kappa}{2} X^e_j \mbox{ where }
X^e_j=X_j(h_e),
\ee
and $X_j(h)$ denotes the right invariant vector fields on $G$ at $h$,
that is $X_j(h):=\mbox{tr}((\tau_j h)^T\partial/\partial h)$.
Thus, the coherent state transform is (following the notation of 
\cite{36a})
\be \label{8.9}
\hat{W}_{\gamma t}:=e^{-\frac{\hat{C}_\gamma}{\hbar}}=
e^{\frac{t}{2}\Delta_\gamma} 
\ee
where we have defined the Laplacian on $\gamma$ by
\be \label{8.10}
\Delta_\gamma=\sum_{e\in E(\gamma)} \Delta_e,\; \Delta_e=\frac{1}{4}
\delta^{ij} X^e_i X^e_j 
\ee
and the heat kernel time parameter has the following interpretation in 
terms of the fundamental constants of the theory
\be \label{8.11}
t:=\frac{\hbar\kappa}{a^{n_D}}\;.
\ee
Notice that $a$ is just a parameter that we have put in by hand to make 
things dimensionless, for instance, it could be $1$cm in quantum general 
relativity in $D+1=4$ spacetime dimensions or $a=10^5$ for 
Yang-Mills in $D+1=4$ and thus is ``large".
The semiclassical limit $\hbar\to 0$ thus corresponds to $t\to 0$.
That $t$ is a tiny positive real number will be crucial in all the estimates 
that we are going to perform in this and the next paper of this series.

The factor of $1/4$ in the definition of $\Delta_e$ relative to 
$(X^e_j)^2$ is due to the factor of $1/2$ in the second Poisson bracket 
of (\ref{8.3}) and it is the same factor which gives $-\Delta_e$ 
the standard spectrum $j(j+1);\;j=0,\frac{1}{2},1,\frac{3}{2},..$ for the 
case of $G=SU(2)$.
 
We can also explicitly compute the quantum operator corresponding to 
$g_e$ in (\ref{8.6}) for arbitrary $G$. We have 
\ba \label{8.12}
\hat{g}_e &=& e^{t\Delta_\gamma/2}\hat{h}_e^{-t\Delta_\gamma/2}
=\sum_{n=0}^\infty\frac{(-t)^n}{2^n n!}[\hat{h}_e,\Delta_e]_{(n)}\nonumber\\
-[\hat{h}_e,\Delta_e] &=& \frac{1}{4}(X^i_e\tau_i\hat{h}_e+\tau_i\hat{h}_e 
X^i_e)=X^i_e\frac{\tau_i}{2}\hat{h}_e-\frac{(\tau_i)^2}{4}\hat{h}_e
\ea
Since $\Delta_\gamma$ commutes with $X^i_e$ we immediately find
\ba \label{8.13}
\hat{g}_e &=&
e^{t\hat{X}^i_e\frac{\tau_i}{4}-t\frac{\tau_i^2}{8}}\hat{h}_e
=e^{-i\hat{p}^j_e\frac{\tau_j}{2}-\frac{t\tau_j^2}{8}}\hat{h}_e
=e^{-i\hat{p}^j_e\frac{\tau_j}{2}} e^{-t\frac{\tau_j^2}{8}}\hat{h}_e
\ea
since $itX^j_e/2=\hat{p}^e_j$
and in the third step we used that the matrix $\tau_j^2$ commutes with 
$\tau_i$. Since the $\hat{p}_j$ are not mutually commuting the exponential 
in (\ref{8.13})
cannot be defined by the spectral theorem, however, we can define 
it through Nelson's analytic vector theorem.
Thus, we find precisely the quantization of 
the polar decomposition (\ref{8.6}) up to a factor of $e^{-\tau_j^2 t/8}$ 
which tends to unity linear in $t\to 0$ as to be expected.
Notice that one obtains
the first line of (\ref{8.12}) from (\ref{8.6}) if one replaces everywhere
$\{.,.\}$ by $[.,.]/(i\hbar)$ and phase space functions by 
operators which holds, of course, by the very construction of the map 
$\hat{W}_t$ \cite{36a}.\\
\\
This accomplishes our goal to write $g_e$ as a function of the $h_e,P_e$
and thus an interpretation of $g_e$ is indeed possible. As we will
discuss all the properties of the corresponding coherent states in 
great detail in \cite{31,32} we will refrain from commenting on them here.
As we will see, these states in fact enjoy all the properties (a)--(h)
that we wanted them to satisfy. In particular, they are diffeomorphism
covariant since, in contrast to \cite{36c}, we have simply managed to 
interprete $\hat{C}_\gamma$ as a function of the 
diffeomorphism covariant functions $h_e,P^e$.

We restrict ourselves here to pointing out that the states constructed there 
will be mainly discussed at the gauge non-invariant and diffeomorphism 
non-invariant level only. There are two good reasons for this restriction.
First of all, both the gauge group and the diffeomorphism group are 
represented unitarily on the Hilbert space \cite{7} and thus expectation 
values of
gauge -- and diffeomorphism invariant operators are in fact gauge -- and
diffeomorphism invariant. It follows that no redundant information is 
produced as far as expectation values are concerned which is enough for
semi-classical considerations. Secondly, while the gauge transformations 
generated by the Hamiltonian constraint are not unitarily represented,
what we can do is to investigate whether the infinitesimal dynamics 
of quantum general relativity as advertized in \cite{17,18} reduces to
that of classical general relatity as $t\to 0$. This would give faith into
the proposal \cite{17,18} and as we will see, the answer is indeed
affirmative \cite{36m}.

More ambitiously, however, one may ask whether it is not possible to
work directly at the gauge -- and diffeomorphism invariant level.
The next two sections outline what can be said about this issue.\\
\\
Remark :\\
The reader may wonder what happens with the quantization ambiguity
labelled by the Immirzi parameter $\beta$ (e.g. \cite{48a}) if one combines 
the quantum
theory with the semi-classical considerations started in this paper.
It is easy to see that the ambiguity, expectedly, does not affect
the classical limit. To see this, recall that the canonical pair is
given by $A_\beta=\Gamma+\beta K,E/(\kappa\beta)$ where $\Gamma$ is the   
spin connection associated with $E$ and $K$ is related to the extrinsic
curvature. Now, for instance, the area of a surface $S$ with normal
co-vector $n_a$ is given by
$$
A(S)=\int_S d^2x\sqrt{E^a_j E^b_j n_a n_b}=
\kappa\beta \int_S d^2x\sqrt{\frac{E^a_j}{\kappa\beta}
\frac{E^b_j}{\kappa\beta} n_a n_b}
$$
and the area operator in the theory with label $\beta$ will be of the form
$\hat{A}_\beta(S)=\beta\hat{A}_1(S)$ where $\hat{A}_1(S)$ has the standard
spectrum of, say \cite{13}. Now the Immirzi parameter also modifies the
classicality parameter $t=\beta\kappa\hbar/a^2$ and the definition
of the momenta $P^e_\beta(E)=P^e_1(E)/\beta$. Consider now a coherent state
peaked at $E$. In the $\beta$-theory the coherent state will then be labelled
by $P^e_\beta(E)$ and the expectation value of the area operator, which
in terms of $\hat{P}^e_\beta$ is of the form
$\hat{A}_\beta(S)
=\beta \sum_e \sqrt{\hat{P}^e_{\beta j}\hat{P}^e_{\beta j}}$, will be by
construction
$<\hat{A}_\beta(S)>=\beta \sum_e \sqrt{P^e_{\beta j}P^e_{\beta j}}
=\sum_e \sqrt{P^e_{1 j}P^e_{1 j}}$, that is, independent of $\beta$.

\section{Coherent States Directly for Gauge Invariant Quantities}
\label{s4}

There are two possibilities for constructing gauge invariant coherent
states. The first possibility consists in group avaraging the 
gauge-variant coherent states of \cite{31} by means of the group
averaging method \cite{7} applied to the gauge group which means
quantizing before reducing. Precisely,
such states will be constructed as
\be \label{4.0}
\Psi^t_{\gamma,\vec{g}}(\vec{h})=
\int_{G^{|V(\gamma)|}} [\prod_{v\in V(\gamma)} d\mu_H(u_v)] \;
\psi^t_{\gamma,\vec{g}}(\{u_{e(0)} h_e u_{e(1)}^{-1}\}_{e\in E(\gamma)})
\ee
where we have assumed that the parameterizations of edges are such
that parameter values $0,1$ respectively correspond to start and end 
respectively. An interesting feature of the state (\ref{4.0})
is that it is separately invariant under gauge transformations of both
$\vec{h},\vec{g}$, a property that is not shared by the diffeomorphism 
group averaged coherent states of the next section. In order to qualify 
as a state on the reduced phase space with respect to the Gauss constraint
one would have to restrict $\vec{g}$, in addition, to the constraint surface
which for the variables $h_e,P_e$ is explicitly described in \cite{37}.
The properties of the states (\ref{4.0}) will be studied in some detail
in \cite{31} so that we can pass on to the second possibility.\\

This second approach to gauge invariant coherent states is the following 
one,  consisting in reducing before quantizing :\\
One directly constructs gauge invariant configuration and momentum
operators on the constraint surface of the Gauss constraint
which leave the space of cylindrical gauge invariant 
functions over a given graph
invariant. Next, one constructs from those new operators with canonical
commutation relations and thus has mapped the problem to that of the
construction of coherent states for the quantization of a particle moving
in a finite number of dimensions for which a natural answer is given
by the usual harmonic oscillator coherent states. \\
We will now outline this idea in some detail :\\
\\
First we must determine suitable, independent, gauge invariant configuration
and momentum operators on a given graph.

Consider a graph $\gamma$ with $E=|E(\gamma)|$ edges and $V=|V(\gamma)|$
vertices. If the gauge group is $N-$dimensional then for each vertex
we have $N$ gauge degrees of freedom which allows us to fix
$NV$ of the $NE$ independent components of the $E$ holonomies $h_e,\;
e\in E(\gamma)$. This reveals that the number of physical configuration
degrees of freedom associated with a graph $\gamma$ is given by
$D(\gamma)=N(E-V)$. (We are considering here generic graphs with
only at least four-valent vertices in order to have non-vanishing
volume; the formula is not correct for the remaining degenerate graphs,
for instance the graph consisting of only a single loop still has
$r$ degrees of freedom while $E=V=2$ with $r$ the rank of the group. We also 
consider only semi-simple Lie groups for definiteness).

Before we construct a suitable set of such $D$ configuration observables,
let us check that the number of gauge invariant momentum observables
also equals $D=N(E-V)$. A suitable set of gauge invariant quantum
operators that can be obtained from electrical field operators alone
consists of a maximal
set of mutually commuting, gauge invariant operators constructed from the
left or right invariant vector fields $^L X^i_e,^R X^i_e$ on the various
copies of the group
associated with the edes $e$ of the graph. Such a choice of invariants
corresponds to the choice of a ``recoupling scheme for the associated
angular momenta''. Let us outline this for $G=SU(2)$ :\\
We can construct
the $E$ Laplacians $\Delta_e=(^R X^i_e)^2$ and for each $n(v)$-valent
vertex $v$ we can construct further mutually commuting $n(v)-3$ invariants
given by the squares of the operators
$(^R X^i_{e_1})+(^R X^i_{e_2}),\;
(^R X^i_{e_1})+(^R X^i_{e_2})+(^R X^i_{e_3}),\;..,
(^R X^i_{e_1})+..(^R X^i_{e_{n(v)-2}})$.
By gauge invariance
$(^R X^i_{e_1})+..(^R X^i_{e_{n(v)}})=0$ so that
$(^R X^i_{e_1})+..(^R X^i_{e_{n(v)-1}})=-(^R X^i_{e_{n(v)}})$ is not another
independent quantity. The choice of these recoupling momenta corresponds to
the choice of a recoupling scheme. Now notice that each edge is connected
to two vertices. Thus the number of recoupling degrees of freedom is
given by
$\sum_{v\in V(\gamma)}(n(v)-3)=2E-3V$ which amounts together with the
$E$ Laplacians to precisely $D=N(E-V)=3(E-V)$ momentum degrees of freedom
as well.\\
In the case of a general group, similar arguments apply.

We now come back to the problem of the construction of
quantum observables with canonical commutation relations from the basic
holonomy and membrane variables $h_e(A)$ and $P^e(A,E)$ respectively.\\
Let us first consider the configuration space operators. Notice that
by the Euler relation \cite{49} there are $L(\gamma)=E-V+1$ generators
(based at an arbitrary but fixed vertex $p$ of $\gamma$)
of the homotopy group $\pi_p(\gamma)$ of $\gamma$.
Thus, choosing a set of such generators
one can construct $D$ independent
configuration degrees of freedom by forming $D$ traces of holonomies
along those loops and their compositions (and products of those if
$r=\mbox{rank}(G)>1$). However, one must be careful
that the ranges of these traces (of products of holonomies along the 
various generators) in the set of real numbers do 
not depend on each other. Let us outline this for $G=SU(2)$ :\\
Choose generators $\alpha_1,..,\alpha_L$ of $\pi(\gamma)$ and define
\be \label{3.33}
t_I=\frac{1}{2}tr(h_{\alpha_I}),\;I=1,..,L
\ee
and since the $\alpha_I$ are independent we have that the $t_I$ take
independently values in $[-1,1]$.

Notice that so far we did not capture any information about the unit
vectors $n_I$ in the representation
$h_{\alpha_I}=t_I 1+\tau_j n_I^j \sqrt{1-t_I^2}$. The scalar products
$n_I^i n_J^i$ are certainly gauge invariant but they cannot be all
independent.
Pick one of the generators, say $\alpha_1$, and decompose the $n_J,\;
J=2,..,L$ into unit vectors parallel and orthogonal $b_J,\; J=2,..,L$ to
$n_1$
\be \label{3.34}
n_J=t_{L+J-1}n_1+\sqrt{1-t_{L+J-1}^2}b_J,\;J=2,..,L
\ee
where the parameters $t_J$ again take independent values in $[-1,1]$.
We can obtain them in terms of traces as
\be \label{3.35}
t_{L+J-1}=
\frac{t_1 t_J-\frac{1}{2}tr(h_{\alpha_1\circ\alpha_J})}
{\sqrt{1-t_1^2}\sqrt{1-t_J^2}}\;.
\ee
Finally, we can also decompose $b_K,\;K=3,..,L$ into unit vectors parallel
and orthogonal $c_K$ to, say, $b_2$
\be \label{3.36}
b_K=t_{2L+K-3}b_2+\sqrt{1-t_{2L+K-3}^2}c_K
\ee
where, of course, $c_K=\epsilon_K c_3, \epsilon_K=\pm 1, K=4,..,L$. Clearly,
$n_1,b_2,c_3$ form an orthonormal basis in $\Rl^3$. In terms of
traces again :
\be \label{3.37}
t_{2L+K-3}=\frac{\frac{t_2 t_K-\frac{1}{2}tr(h_{\alpha_2\circ\alpha_K})}
{\sqrt{1-t_2^2}\sqrt{1-t_K^2}}-t_{L+1}t_{L+K-1}}
{\sqrt{1-t_{L+1}^2}\sqrt{1-t_{L+K-1}^2}}\;.
\ee
Similarly, we could also express the $L-3$ dscrete variables $\epsilon_M,
L=4,..,M$ in terms of traces along the lines given above but we will not
display the explicit formulae here. Rather, by means of the following trick
we can get rid of them : define $t'_{2L+K-3}:=t_{2L+K-3}$ and new
parameters $t_{2L+K-3},\;K=4,..,L$ by
\be \label{3.38}
t'_{2L+K-3}=2t_{2L+K-3}^2-1 \mbox{ and } \epsilon_K\sqrt{1-(t'_{2L+K-3})^2}
=2\sqrt{1-t_{2L+K-3}^2}t_{2L+K-3}
\ee
with, again, $t_{2L+K-3}\in [-1,1]$.

Obviously, the above equations (\ref{3.33})-(\ref{3.38}) define precisely
$3(L-1)=3(E-V)$ continuous gauge invariant parameters $t_I, I=1,..,D$
with independent range in $[-1,1]$.
The map between the selected traces
and these variables is singular but 
the subset of the space $[-1,1]^D$ where this map is singular
is of Lebesgue measure zero and thus is irrelevant for $L_2$ functions.
In any case, all traces of loops on $\gamma$ can be written
as definite functions of the $D$ variables $t_I$ since any such function
is a polynomial in the quantities $n_I^i n_J^i$ and we just need to
substitute (\ref{3.34}), (\ref{3.36}).

From now on we will assume that we have constructed precisely $D(\gamma)$
independent, gauge invariant configuration variables $t_I,\; I=1,,D$
for every graph $\gamma$ with range in $[-1,1]$ along lines similar as
above. This suggests the following strategy :\\
We would like to map the problem at hand to the problem of $D$ uncoupled
harmonic oscillators. We achieve this by defining new variables
\be \label{3.39}
x_I:=\mbox{arctanh}(t_I)=\frac{1}{2}\ln(\frac{1+t_I}{1+t_I})
\;\Leftrightarrow\;t_I=\mbox{tanh}(x_I)
\ee
which take values in the whole real line. We can now consider the
Hilbert space ${\cal H}_\gamma=L_2(\Rl^D,d^Dx)$ and construct the usual
coherent states associated with the annihilation operators
$\hat{z}_I=\hat{x}_I+i\frac{t}{\hbar}\hat{p}_I$ where $\hat{p}_I=-i\hbar
\partial/\partial x_I$ and $t$ is a dimensionless parameter.

This is, however, not the end of the story. Namely, in order to interprete
these coherent states in terms of the original quantities, we must make the
connection with the classical theory.
For the configuration variables the interpretation is obvious
through the formulae (\ref{3.33})-(\ref{3.37}). For the momentum variables
this is less obvious. The way to proceed is to first express the
operators $\hat{p}_I$ in terms of right invariant vector fields on
functions cylindrical with respect to $\gamma$ and then to express the latter
in terms of the phase space variables. In order to do that we write
\be \label{3.40}
\partial_{x_I}=\frac{\partial t_J}{\partial x_I}
\frac{\partial\theta_{e,i}}{\partial t_J} \partial_{\theta_{e,i}}
\ee
where $\theta_{e,i}=\theta_{e,i}(t_I,t_\mu),\;e\in E(\gamma),\;
i=1,..,N,\; \mu=1,..,NE-D$ are the $NE$ angle parameters which coordinatize
the $E$ copies of $G$ and which we can think of as functions of the
$t_I$ and remaining gauge degrees of freedom $t_\mu$. Now, there
exists a map
\be \label{3.41}                                 
\partial_{\theta_{e,i}}=F_{ij}(\theta_{e,k})(^R X^j_e)
\ee
which generically (that is, almost everywhere) is also non-singular
and which allows us to write (\ref{3.40}) in the form
\be \label{3.42}
\partial_{x_I}=F_{I,ei}(\{h_{e'}\}_{e'\in E(\gamma}) (^R X^j_e)\;.
\ee
The final step consists in expressing the right invariant vector fields
in terms of electric fields in the form of membrane operators
which has been done in section \ref{s2.1} where they have been
called $P^e_j$.

We can then finally think of $\partial_{x_I}$ as a definite function
of the $\{\hat{h}_e,\hat{P}^e_i\}_{e\in E(\gamma)}$
with an obvious classical limit. Of course, the formula (\ref{3.42})
is far from simple.

Notice that in the course of the construction we have defined a new
Hilbert space ${\cal H}_\gamma=L_2(\Rl^{D(\gamma)},d^{D(\gamma)}x)$
which, however, is unitarily equivalent to the projection of
the kinematical Hilbert space $\cal H$ of section \ref{s2} to the
space of functions cylindrical over $\gamma$ (after integrating out gauge
degrees of freedom) which also shows that these Hilbert spaces are
cylindrically consistent so that they line up to a big Hilbert space
in the projective limit, unitarily equivalent to $\cal H$.\\
\\
We close this section with a number of comments :\\
(i)\\
The advantage of this approach as compared to the one outlined in the
previous section is that we are guaranteed to fulfill all the requirements
(a)-(h) without going through considerable amount of functional
analytic work since we can just copy all the results known from
the harmonic oscillator coherent states.\\
(ii)\\
A disadvantage is that the coherent states so constructed in terms
of the $x_I$ are not easily expressed in terms of the gauge invariant
spin-network functions in terms of which the spectra of important
operators, such as the geometrical ones \cite{11,12,13,14,15,16},
are well known.\\
(iii)\\
Finally, the reader may ask why we did not work at the gauge non-invariant
level to begin with, obtain harmonic oscillator kind of  
coherent states for the gauge-variant quantities and only then solve the 
Gauss constraint. 
While this would simplify the analysis considerably since all the
gauge angles $\theta^j_e$ could be taken as independent configuration 
variables
and we could relate the conjugate derivative operators much more
easily to the right invariant vector fields, unfortunately
the gauge invariant subspace of the coherent states constructed from
the gauge non-invariant quantities $\theta^j_e$ is not explictly known.
The only known procedure is to write them in terms of non-gauge invariant
spin-network functions and then to keep only the gauge invariant 
combinations (this can be done alternatively by integrating those states 
over the gauge degrees of freedom as in (\ref{4.0})). 
However, the coherent states are an infinite
superposition of harmonic oscillator eigenstates each of which is an
infinite superposition of spin-network states (in the $L_2$ sense) because
the relation between the $\theta_{e,I}$ and the $h_e$ are not at all
polynomial.
Thus, the amount of work to be done to solve the Gauss constraint is
considerably larger, if possible at all, than to define gauge invariant
coherent states directly. \\
iv)\\
Finally, the complications mentioned in ii) of course also apply
if one works entirely with gauge variant variables $\theta^j_e$ 
mentioned in iii) without caring about the Gauss constraint,
the only simplication as compared to iii) is that the construction of the 
$t_I$ is not necessary.\\
\\
\\
To summarize, the coherent states defined in sections \ref{s3.2}, \ref{s3.3}
may reveal the required properties (a)-(h) less obviously, on the other
hand, the operators that appear in applications have a much simpler
action on these than on the ones that were constructed in the present 
section. Thus altogether, at least for analytical purposes the set of states 
of section \ref{s3.3} seems to be preferred.

\section{Diffeomorphism Invariant Coherent States}
\label{s5}

Given a coherent state $\psi^t_{\gamma,Z}$
we can group average it with respect to the diffeomorphism constraint
\cite{7} and obtain (we discard certain technicalities 
that come from graph symmetry factors, see \cite{23}, whose notation 
we follow, for details) 
\be \label{5.1}
\eta_{Diff}\cdot\psi^t_{\gamma,\vec{g}}=\sum_{\lambda,n}e^{-t\lambda}
T_{\gamma,\lambda,n}(\vec{g})[T_{\gamma,\lambda,n}] \mbox{ and }
\eta_{Diff}\cdot\Psi^t_{\gamma,\vec{g}}=\sum_{\gamma'\in\gamma}
\eta_{Diff}\cdot\psi^t_{\gamma,\vec{g}}
\ee
where $[\psi]$ denotes the orbit of the state $\psi$ under $Diff(\Sigma)$,
typically
\be \label{5.2}
[T_{\gamma,\lambda,n}]=\sum_{\gamma'\in [\gamma]}T_{\gamma',\lambda,n} \;.
\ee
where $[\gamma]$ denotes the orbit of $\gamma$. Here, as in section
\ref{s3.1} we have written coherent states in terms of eigenfunctions 
$T_{\gamma,\lambda,n}$ of a general complexifier with eigenvalue 
$\lambda$ and degeneracy level $n$
each of which can be decomposed in terms of spin-network
functions with non-trivial dependence on every edge of that graph. 
This requirement is very important in order for group averaging to 
be well-defined and thus excludes, in particular, the possibility
to average infinite graphs as we will see in \cite{33}, at least not
without some kind of renormlization as discussed there, see 
\cite{54} for a general discussion.

If $\hat{C}$ is diffeomorphism invariant, as it is the case 
for $\hat{v}$ above then $\lambda$ is a diffeomorphism invariant quantity.

Although the state (\ref{5.1}) is certainly diffeomorphism invariant,
being a linear combination of diffeomorphism invariant states,
it depends not only on $[\gamma]$ and the equivalence class of complex 
holonomies under diffeomorphisms
$[\vec{g}]$, but explicitly on the representants. In other words, while
under diffeomorphisms also $g_e\to \varphi\cdot g_e=g_{\varphi^{-1}(e)}$, 
(\ref{5.2})
is not invariant under mapping $\vec{g}$ by a diffeomorphism in contrast
to what happened in (\ref{4.0}) with respect to the gauge group.
This is unsatisfactory because
on the diffeomorphism invariant Hilbert space ${\cal H}_{Diff}$, which is 
the Cauchy completion
of states of the form $\eta_{Diff}\cdot f,\; f\in Cyl$ under the inner 
product
\be \label{5.3}
<\eta_{Diff}\cdot f,\eta_{Diff}\cdot g>_{Diff}:=
[\eta_{Diff}\cdot f](g)
\ee
where the latter denotes the application of the distribution 
$\eta_{Diff}\cdot f$ to the test function $g$, the inner product between
diffeomorphism invariant coherent states should depend only on $[\vec{g}],
[\vec{g}']$ and not on the representants. In particular, this leads 
to the following problem :
Suppose that $(A,E)$ and $(A',E')$ are diffeomorphic points of the 
classical phase space and compute from these $g_e=g_e(A,E)$ as in
section \ref{s3.1} or \ref{s3.2} and similar for the primed quantities.
Then if these quantities differ in the range of $\gamma$ then the inner 
product
\be \label{5.4}
<\eta_{Diff}\cdot\psi^t_{\gamma,\vec{g}},
\eta_{Diff}\cdot\psi^t_{\gamma,\vec{g}'}>_{Diff}
=\sum_{\lambda,n}e^{-2t\lambda}\overline{T_{\gamma,\lambda,n}(\vec{g})}
T_{\gamma,\lambda,n}(\vec{g}')
\ee
will be small by the very definition of a coherent state, that is, these
states are almost orthogonal with respect to $<.,.>_{Diff}$. This is
certainly not what we want.

The reason for this is, of course, that there are too many of the states
$\eta_{Diff}\cdot\Psi^t_{\gamma,Z}$. We should identify
all those that are labelled by those $\vec{g}'$ which lie in the same 
equivalence class under diffeomorphisms as $\vec{g}$.
This can be done by choosing a representant $Z_0([Z])$ in every
equivalence class $[Z]$ where $Z$, as before, stands for phase space
points $(A,E)$ or an actual complex connection depending on whether
we choose coherent states based on otion 2) or 1). Notice that this is not, 
in general, equivalent
to fixing a gauge because choosing a representant is possible also if there
does not exist a global gauge fixing condition as it is typically the case in
field theories.

One might think that one could alternatively define diffeomorphism invariant
coherent states by heat kernel evolution, followed by analytical
continuation, of the $\delta$ distribution with respect to
$<.,.>_{Diff}$ given by (notice that
$T_{[\gamma],\lambda,n}(A)=T_{[\gamma],\lambda,n}([A])$)
\be \label{5.5}
\delta_{[\gamma],[A]}([A']):=\sum_{\lambda,n}T_{[\gamma],\lambda,n}(A)
\overline{T_{[\gamma],\lambda,n}(A')},
\ee
however, the resulting state
\be \label{5.6}
\psi^t_{[\gamma],[Z]}([A])=\sum_{\lambda,n}e^{-t\lambda}
T_{[\gamma],\lambda,n}(Z)\overline{T_{[\gamma],\lambda,n}(A)},
\ee
is no longer normalizable with respect to
$<.,.>_{Diff}$ so that we are forced to adopt the above strategy.

To summarize, we pick arbitrary but fixed representant functions
\ba \label{5.7}
\gamma_0 &:& [\Gamma^\omega_0]\mapsto\Gamma^\omega_0;\;[\gamma]\mapsto
\gamma_0([\gamma])
\mbox{ and }\nonumber\\
Z_0 &:&\; M_{Diff}\mapsto M;\;
[Z]\mapsto Z_0([Z])
\ea
from the sets of equivalence classes under diffeomorphisms of
piecewise analytical graphs and
from the phase space $M_{Diff}$ reduced with respect to the diffeomorphism
constraint to the full phase space $M$ respectively
and we define diffeomorphism invariant coherent states by
\be \label{5.8}
\Psi_{[\gamma],[Z]}^{t,Z_0,\gamma_0}:=
\eta_{Diff}\cdot \Psi^t_{\gamma_0([\gamma]),Z_0([Z])}\;.
\ee
The function $\gamma_0$ is necessary on top of $Z_0$ since 
a coherent state on $\gamma$ depends on $Z$ only at $\gamma$ and not 
everywhere.
The inner product between these states is given through
\be \label{5.9}
<\psi_{[\gamma],[Z]}^{t,Z_0,\gamma_0},
\psi_{[\gamma'],[Z']}^{t,Z_0,\gamma_0}>_{Diff}=
<\psi^t_{\gamma_0([\gamma]),Z_0([Z])},\psi^t_{\gamma_0([\gamma']),Z_0[Z']}>
\ee
using the orthogonality of the $\psi^t_{\gamma,Z}$ for different $\gamma$. 
Notice that in the last line we just have the kinematical inner product on
$\cal H$. It follows from (\ref{5.9}) immediately that the
diffeomorphism invariant coherent states so defined are localized in the
same way as the kinematical ones are. The Ehrenfest properties 
cannot be verified because we would need a complete set of observables
on the Hilbert space ${\cal H}_{Diff}$ but it is sufficient to know that
these states are peaked on $[Z]$ for every $[\gamma]$ in order to make
semi-classicl approximations.  Moreover, it also follows from
(\ref{3.9}) that the group average 
of the projective limit $\Psi^t_Z$ coherent state is normalizable
with respect to $<.,.>_{Diff}$ if and only if $\Psi^t_Z$ is normalizable
with respect to $<.,.>$.

As we have explicitly indicated in (\ref{5.8}), the coherent states depend
on the representant functions (\ref{5.7}). 
But
\be \label{5.3a}
\eta_{Diff}\cdot\Psi^t_{\gamma,Z}=
\eta_{Diff}\cdot\Psi^t_{\gamma_0([\gamma]),\varphi_0^\ast Z}
\ee
where $\varphi_0(\gamma_0([\gamma]))=\gamma$ and $\varphi_0^\ast$ is the 
action of diffeomorphisms on phase space points $Z$. 
Thus, it is only the relation between $\gamma_0$ and $Z_0$ which makes a
difference (has a dffeomorphism invariant meaning) because in the pair
$\gamma_0',Z_0'$ we can always replace
$\gamma_0'$ by $\gamma_0$ at the price of changing $Z_0'$. In other words,
if we fix $\gamma_0$ once and for all as we can without loss of generality,
then our choice of diffeomorphism invariant coherent states is entirely
labelled by $Z_0$. This choice is to be interpreted as a choice of basis of
diffeomorphism invariant coherent states. The inner products between
members of different bases have no definite locality properties as we have
shown in (\ref{5.4}). But this is in general true for different sets of
coherent states even in systems with only a finite number of degrees of
freedom. After all, the requirement of localization does not determine
a coherent state uniquely, not even up to unitary equivalence because
all that is required is that the inner product between such states is unity
if their labels coincide and is ``small" otherwise where the notion of
smallness depends on the basis.\\
Thus the dependence of the states on $Z_0$ is not a bad but in fact
an expected property.

Notice further that some of these diffeomorphism invariant coherent states
also lie in the kernel of the Hamiltonian constraint operator
defined in \cite{17,18,23} : we just have to choose $[\gamma]$
in such a way that the range of the Hamiltonian constraint in the set of
linear combinations of spin-network functions cannot contain a spin-network 
state
whose underlying graph lies in the class $[\gamma]$. As shown in
\cite{23}, there are an infinite number of such states. This observation
may be a starting point for the construction of semiclassical states which
lie in the kernel of all three types of constraints : the Gauss-,
Diffeomorphism- and Hamiltonian constraint.

\section{Model for Gauge Invariant Coherent States : Euclidean 2+1 gravity}
\label{s6}

As we have mentioned in section \ref{s3.1}, the volume operator
qualifies best as a complexifier in $D=2$. 
For Euclidean 2+1 gravity we have $D=2$ and $G=SU(2)$. The volume operator
in two dimensions was derived in \cite{19}. The spectrum of that
operator for at most three-valent vertices was also computed analytically
there.
In this section we focus on a Hilbert space for this theory which
has finite linear combinations of spin-network states on at most
three-valent graphs as a dense subset because otherwise the spectrum
is only known numerically.

There are two cases to consider : either (A) no two of the three edges
$e_1,e_2,e_3$ meeting at
a vertex are co-linear or (B) there is a co-linear pair,
say $e_1,e_2$ (the third case, that
all three edges are co-linear is excluded because the volume would vanish).
Let $\vec{j}=(j_1,j_2,j_3)$ be the spins with which the three edges are
coloured (for at most three-valent graphs the space of vertex-contractors
is one dimensional and thus $J_v=1$ is suppressed in what follows;
this is also the reason why these spin-network states are eigenstates of the
volume operator).
\\
The square of the eigenvalues of the volume operator for a given vertex
in the two cases are \cite{19}
\ba \label{6.1}
\lambda_v(\vec{j})&=&\frac{9}{4}
[2(\Delta_1\Delta_2+\Delta_2\Delta_3+\Delta_3\Delta_1)
-(\Delta_1^2+\Delta_2^2+\Delta_3^2)]-\frac{1}{2}[\Delta_1+\Delta_2+\Delta_3]
\mbox{ (A)}\nonumber\\
\lambda_v(\vec{j})&=&
[2(\Delta_1\Delta_2+\Delta_2\Delta_3+\Delta_3\Delta_1)
-(\Delta_1^2+\Delta_2^2+\Delta_3^2)]-\Delta_3
\mbox{ (B)}
\ea
where $\Delta_I=-j_I(j_I+1)$. At first sight it seems that in this case we
can eve take $n=1$ since the $\Delta$'s appear squared in leading order
which would be sufficient to make the series of the coherent state
converge.
However, this is not the case : For instance we can consider the case
that $j_1=const.$ and $j_2\to\infty$. Then, due to gauge invariance $j_3$
is of the same order as $j_2$ and therefore the leading order of the
square bracket in (\ref{6.1}) is only $j_2^2$. We choose $n=2$ in what
follows. It is then easy to see that in this case the complex connection is 
explicitly given by $Z_a^j=A_a^j-i f_a^j$ where 
$$
f_a^j\propto \frac{V}{\sqrt{\det(q)}}\epsilon_{jkl}\epsilon_{ab}
(\epsilon_{kmn}\epsilon_{cd} E^c_m E^d_n) E^b_l
$$

We will for our example analyze only the simplest non-trivial graph, a kink 
(or double kink) $\alpha$ with two edges and one (or two) two-valent vertices. 
This corresponds
to, say $j_2=0$, in (\ref{6.1}) and $j:=j_1=j_3$. Then we obtain the simple
eigenvalue  $\lambda_j=\lambda_v^2=-\Delta=j(j+1)$, in other words, on this
graph
the volume operator reduces to (two times) the square root of the Laplacian
on the copy of the group corresponding to $h:=h_\alpha,\;
\alpha=e_1\circ e_3^{-1}$. A complete orthonormal basis of gauge invariant
spin-network functions
is given by the characters
$\chi_n(h)=tr(\pi_{n/2}(h)),\;n=0,1,2,..$.

On the kink, the coherent state is simply given by 
\be \label{6.2}
\Psi_{g,\alpha,t}(h)=\sum_{n=0}^\infty e^{-\frac{t}{4}\lambda_n}\chi_n(g)
\overline{\chi_n(h)}
\ee
where $g=h_\alpha(Z),\; \lambda_n=n(n+2)=4(\lambda_j)_{n=2j}$. The series
(\ref{6.2})
converges for any $g\in SL(2,\Cl)$ as shown in \cite{36b}.

We wish to show that the state (\ref{6.2}) diagonalizes the gauge invariant
operator
\be \label{6.3}
\hat{T}^\Cl:=\hat{W}_t\hat{T}(\hat{W}_T)^{-1},\,\hat{W}_t=e^{t\Delta},
\;\hat{T}=
tr(\hat{h})\;.
\ee
Denoting $T_n=tr(h^n),\;n=0,1,..,\;T=T_1$ we notice the identity
\be \label{6.4}
\chi_n=\left\{ \begin{array}{cc}
1+T_2+T_4+..+T_N & \mbox{ : n even}\\
T_1+T_3+T_5+..+T_N & \mbox{ : n odd}
\end{array} \right.
\ee
from which follows that $T\chi_n=\chi_{n+1}+\chi_{n-1}$ by using the
$SU(2)$ Mandelstam identity $T T_n=T_{n+1}+T_{n-1}$. It is understood that
$T_{-1}=0$. Let $T^\Cl=tr(g)$, then
\ba \label{6.5}
&&(\hat{T}^\Cl\Psi_{g,t})(h)= e^{t\Delta}\sum_n \chi_n(g) T\chi_n(h)
= e^{t\Delta}\sum_n \chi_n(g) [\chi_{n+1}(h)+\chi_{n-1}(h)]
\nonumber\\
&=& \sum_n \chi_n(g) [e^{-t\lambda_{n+1}}\chi_{n+1}(h)
+e^{-t\lambda_{n-1}}\chi_{n-1}(h)]\nonumber\\
&=& \sum_n [\chi_{n+1}(g)+\chi_{n-1}(g) e^{-t\lambda_n}\chi_n(h)]
=T^\Cl \psi_g(h)\;.
\ea
From this and the general discussion in section \ref{s3} it easily follows
that $\hat{T}^\Cl$ and $(\hat{T}^\Cl)^\dagger$
respectively have expectation values $T^\Cl$ and $\overline{T^\Cl}$
respectively, moreover, we have the uncertainty relation
\be \label{6.6}
<(\Delta\hat{x})^2><(\Delta\hat{y})^2>\ge\frac{|<[\hat{x},\hat{y}]>|^2}{4},
\ee
with $\hat{x}=\frac{1}{2}(\hat{T}^\Cl+(\hat{T}^\Cl)^\dagger),\;
\hat{y}=\frac{1}{2i}(\hat{T}^\Cl-(\hat{T}^\Cl)^\dagger)$.

The inner product between two coherent states is given by (we suppress
the label $\alpha$ in what follows)
\be \label{6.7}
<\Psi^t_g,\Psi^t_{g'}>
=\sum_n e^{-2t\lambda_n}\overline{\chi_n(g)}
\chi_n(g')
\ee
We will now show that the overlap integral
\be \label{6.8}
I^t(g,g'):=\frac{|<\Psi^t_g,\Psi^t_{g'}>|^2}
{<\Psi^t_g,\Psi^t_g><\Psi^t_{g'},\Psi^t_{g'}>}
\ee
decays exponentially fast with $|tr(g)-tr(g^\Cl)|$ as $t\to 0$, i.e.
in the classical limit $\hbar\to 0$. The proof uses the
Euler-MacLaurin estimate for the difference between a series and its
replacement by an integral \cite{52} which turns out to vanish in our
limit $t\to 0$.

To begin with, recall that the characters are explicitly given by
\be \label{6.9}
\chi_n(h)=\frac{\sin((n+1)\phi)}{\sin(\phi)} \mbox{ where }
2\cos(\phi):=tr(h),\;\phi\in[0,\pi]
\ee
for any $h\in SU(2)$. Formula (\ref{6.9}) is entire analytic in
$\phi$ and is readily extended to $g\in SL(2,\Cl)=SU(2)^\Cl$
\be \label{6.10}
\chi_n(g)=\frac{\sin((n+1)\theta)}{\sin(\theta)} \mbox{ where }
2\cos(\theta):=tr(g),\;\theta=\phi-is,\;\phi\in[0,\pi],\;s\in\Rl\;.
\ee
Since the characters are class functions, we can always rotate $g$
into a maximal torus of $SU(2)$ and so we can think of $g$ as given
by $g=\exp(\theta\tau_3)$. Notice that the Weyl subgroup acts on the torus
by $\theta\to-\theta$ and indeed (\ref{6.9}), (\ref{6.10}) are still invariant
under it. We can therefore restrict, without loss of generality to
$s\in [0,\infty]$. In general, $\theta=\pm\sqrt{(\theta_i)^2},\;
g=\exp(\theta_i\tau_i)$.

We now compute
\ba \label{6.11}
<\Psi_{g,t},\Psi_{g',t}> &=&\frac{1}{\sin(\bar{\theta})\sin(\theta')}
\sum_{n=0}^\infty e^{-\frac{t}{2}[(n+1)^2-1]}\sin((n+1)\bar{\theta})
\sin((n+1)\theta') \nonumber\\
&=&\frac{e^{t/2}}{\sin(\bar{\theta})\sin(\theta')}
\sum_{n=1}^\infty e^{-\frac{t}{2}n^2}\sin(n\bar{\theta})
\sin(n\theta') \nonumber\\
&=&-\frac{e^{t/2}}{4\sin(\bar{\theta})\sin(\theta')}
\sum_{n=1}^\infty e^{-\frac{t}{2}n^2}\times\nonumber\\
&&\times [\exp(in[\bar{\theta}+\theta'])+\exp(-in[\bar{\theta}+\theta'])
-\exp(in[\bar{\theta}-\theta'])-\exp(-in[\bar{\theta}-\theta'])]
\nonumber\\
&=&-\frac{e^{t/2}}{4\sin(\bar{\theta})\sin(\theta')}
\sum_{n\in Z-\{0\}} e^{-\frac{t}{2}n^2}
[\exp(in[\bar{\theta}+\theta'])-\exp(in[\bar{\theta}-\theta'])]
\nonumber\\
&=&-\frac{e^{t/2}}{4\sin(\bar{\theta})\sin(\theta')}
\sum_{n\in Z} e^{-\frac{t}{2}n^2}
[\exp(in[\bar{\theta}+\theta'])-\exp(in[\bar{\theta}-\theta'])]
\ea
where in the last step we have noticed that the term $n=0$ vanishes.
Let now $x_n:=\sqrt{t}n,\;\Delta x:=x_{n+1}-x_n=\sqrt{t}$, then
(\ref{6.11}) can be written in the form
\be \label{6.12}
<\Psi_{g,t},\Psi_{g',t}>=
-\frac{e^{t/2}}{4\sqrt{t}\sin(\bar{\theta})\sin(\theta')}
\sum_{n\in Z} \Delta x\; e^{-\frac{t}{2}x_n^2}
[\exp(ix_n\frac{\bar{\theta}+\theta'}{\sqrt{t}})
-\exp(ix_n\frac{\bar{\theta}-\theta'}{\sqrt{t}})]
\ee
which suggests to replace the sum by a Riemann integral for small $t$.
It would be literally a Riemann sum if it was not for the explicit
$t$-dependence of the integrand. Thus, the following expression
is only an approximation to (\ref{6.12}) which becomes exact as
$t\to 0$
\ba\label{6.13}
i^t(g,g') &=&
-\frac{e^{t/2}}{4\sqrt{t}\sin(\bar{\theta})\sin(\theta')}
\int_{-\infty}^\infty dx\; e^{-\frac{t}{2}x^2}
[\exp(ix\frac{\bar{\theta}+\theta'}{\sqrt{t}})
-\exp(ix\frac{\bar{\theta}-\theta'}{\sqrt{t}})]
\nonumber\\
&=&
-\frac{e^{t/2}}{4\sqrt{2\pi t}\sin(\bar{\theta})\sin(\theta')}
[\exp(-\frac{(\bar{\theta}+\theta')^2}{2t})
-\exp(-\frac{(\bar{\theta}-\theta')^2}{2t})]
\ea
where we have used a Cauchy integral formula. The overlap integral
thus is approximated by
\ba \label{6.14}
\tilde{I}(g,g',t)&=&\frac{|i^t(g,g')|^2}{i^t(g,g)i^t(g',g')}
\nonumber\\
&=&\frac{|\exp(-\frac{(\bar{\theta}+\theta')^2}{2t})
-\exp(-\frac{(\bar{\theta}-\theta')^2}{2t})|^2}
{[\exp(-\frac{(\bar{\theta}+\theta)^2}{2t})
-\exp(-\frac{(\bar{\theta}-\theta)^2}{2t})]
[\exp(-\frac{(\bar{\theta}'+\theta')^2}{2t})
-\exp(-\frac{(\bar{\theta}'-\theta')^2}{2t})]},
\ea
or when decomposing $\theta=\phi+is,\theta'=\phi'+is'$
\be \label{6.15}
\tilde{I}(g,g',t)
=\frac{|\exp(-\frac{([\phi+\phi']-i[s-s'])^2}{2t})
-\exp(-\frac{([\phi-\phi']-i[s+s'])^2}{2t})|^2}
{[\exp(-2\frac{\phi^2}{t})-\exp(2\frac{s^2}{t})]
[\exp(-2\frac{(\phi')^2}{t})-\exp(2\frac{(s')^2}{t})]}\;
\ee
We now multiply numerator and denominator of (\ref{6.15}) with
$\exp([-s^2-(s')^2+\phi^2+(\phi')^2])/t$ and obtain
\ba \label{6.16}
\tilde{I}(g,g',t)
&=&\frac{|\exp(-\frac{[\phi\phi'+ss']-i[(\phi+\phi')(s-s')]}{t})
-\exp(\frac{[\phi\phi'+ss']+i[(\phi-\phi')(s+s')]}{t})|^2}
{4\mbox{sinh}(\frac{\phi^2+s^2}{t})
\mbox{sinh}(\frac{(\phi')^2+(s')^2}{t})}
\nonumber\\
&=&\frac{\mbox{cosh}(2\frac{\phi\phi'+ss'}{t})
-\mbox{cos}(2\frac{\phi s'-\phi' s}{t})}
{2\mbox{sinh}(\frac{\phi^2+s^2}{t})
\mbox{sinh}(\frac{(\phi')^2+(s')^2}{t})}\;.
\ea
Now, for $\theta,\theta'\not=0$, in the limit $t\to 0$ we have (since
cos is a bounded function and $ss',\phi\phi'\ge 0$)
\be \label{6.17}
\tilde{I}(g,g',t)\to\exp(-\frac{[\phi-\phi']^2+[s-s']^2}{t})
\ee
which is indeed rapidly vanishing as $t\to 0$ unless $\theta=\theta'$
in which case it equals unity as it should.

If either of $\theta,\theta'$ vanishes, say $s'=\phi'=\theta'=0$ then
expression (\ref{6.16}) is of the type $0/0$ and we can evaluate it,
provided the limit exists, by picking up the leading order terms
of numerator and denominator by Cauchy's formula. It turns out to
be sufficient to keep the terms of second order in $s',\phi'$.
The numerator becomes
$$
\frac{1}{2}[(2\frac{\phi\phi'+ss'}{t})^2+(2\frac{\phi s'-\phi' s}{t})^2]
+O((\theta')^3=2\frac{(\phi^2+s^2)((\phi')^2+(s')^2)}{t^2}+O((\theta')^3)
$$
while the denominator becomes
$$
2\mbox{sinh}(\frac{s^2+\phi^2}{t})\frac{(\phi')^2+(s')^2}{t}+..
$$
where the dots denote terms of at least fourth order in $s',\phi'$.
Thus, (\ref{6.16}) has the well-defined limit
\be \label{6.18}
\tilde{I}(g,g'=1,t)=\frac{s^2+\phi^2}{t\mbox{sinh}(\frac{s^2+\phi^2}{t})}
\ee
which is again exponentially damped as $t\to 0$ unless $s=\phi=0$ in which
case it equals unity as it should.

To conclude, the overlap integral $I^t(g,g')\to_{t\to 0}
\tilde{I}(g,g',t)$ is exponentially damped unless $T^\Cl=(T')^\Cl$.

Next, we should also show that the normalized coherent state itself, in 
both the configuration and the momentum representation, is peaked. 
These and other issues will be much more systematically analyzed for 
general graphs in \cite{31} by using the Poisson summation formula. \\
\\
\\
\\
{\large Acknowledgements}\\
\\
We thank the Institute of Theoretical Physics at Santa Barbara for 
hospitality where part of this work has been completed. Special thanks
to the organizers of the workshop ``Classical and Quantum Physics of Strong
Gravitational Fields'', Santa Barbara, January -- June 1999, for 
establishing an inspriring research environment at the ITP. 
We also thank A. Ashtekar, L. Bombelli, R. Gambini, J. Lewandowski,
R. Loll and J. Pullin for for general discussions about semi-classical 
quantum general relativity, A. Ashtekar and L. Bombelli for explaining  
the idea of random weaves to us and O. Winkler for a careful reading of the 
manuscript.

This research project was supported in part by the 
National Science Foundation of the USA under grant PHY94-07194 to the 
ITP, Santa Barbara.

\end{document}